\documentclass[preprint]{aa}
\usepackage{graphicx}
\usepackage{txfonts}                                       
\usepackage{natbib}

\begin{document}
\title{Evolution of stellar-gaseous disks in cosmological haloes}

\author{Anna Curir,\inst{1}
 Paola Mazzei\inst{2}
  and Giuseppe Murante\inst{1}}
\offprints{ A. Curir}
\institute{INAF-Osservatorio Astronomico di Torino. Strada Osservatorio 20
 -10025 Pino Torinese (Torino). Italy.   e-mail: curir@oato.inaf.it
\and 
     INAF-Osservatorio Astronomico di Padova. Vicolo Osservatorio 5 - 35122
 Padova. Italy. e-mail: paola.mazzei@oapd.inaf.it
}
\date{Received/Accepted}


\abstract
{}
{We explore the growth and the evolution of the bar instability 
in stellar-gaseous disks embedded 
in a suitable dark matter halo evolving in a fully consistent cosmological
framework.  The aim of this paper is to point out the impact of different gas
fractions on the bar formation, inside disks of different disk-to-halo mass 
ratio, and the role of the cosmological framework.}
{We perform  cosmological simulations with the same  disk-to-halo mass ratios
as in a previous work where the gas was not taken into account. We 
compare results of the new simulations with the previous ones to
investigate the effect of the gas by analysing the morphology of the
stellar and gaseous components,
the stellar bar strength and the behaviour of its pattern speed.}
{In our cosmological simulations, inside  dark-matter dominated disks,  a 
stellar bar, lasting 10 Gyr,  is still living at $z=0$ even if the gaseous fraction 
exceeds half of the disk mass. However, in the most massive disks we find a 
threshold value (0.2) of the gas fraction able to destroy the bar.\\
The stellar bar strength is enhanced by the gas and  in the  more massive
disks higher gas fractions increase
the bar pattern speed.} 
{}


\titlerunning{Bars in stellar-gaseous disks}
\authorrunning{Curir,  Mazzei \& Murante}
\maketitle

\section{Introduction}

In a pioneering paper  \citet{ber98} presented the numerical
results describing the impact of a gas fraction inside a stellar disk on the
stellar orbits and on the whole bar instability. Their model galaxy containing stars and gas was
compared with a pure stellar model, subject to the same initial conditions,
chosen to produce the development of a strong stellar bar.

They showed that dynamical instabilities become milder in the presence of 
the gas component, and therefore a weakening of the bar itself is observed as
a gradual process.   Moreover, the overall
evolution is accelerated because of the larger central mass concentration and
the resulting decrease in the characteristic dynamical time.
 
The gas behaviour in disk galaxies and its connections with the bar feature was
studied in several papers  (
\citet{friebe93}; \citet{fux99}; \citet{ber01}; \citet{atha02}; 
\citet{bot03}; \citet{Bou05}, \citet{MiWo}, \citet{Stinson}).
However in all these works 
the evolution of the disk or of the disk+halo simulated system arises in a isolated framework,
outside the cosmological scenario, even if  progressive efforts
for improving the model, accounting for more recent knowledges coming from 
the cosmological 
hierarchical clustering scenario of structures formation about  
density distribution and  concentration of the dark matter (DM) halos,
have been performed  in
the last years (\citet{Cu99}; \citet{Ma01}; \citet{athami02}).
Here, for the first time, we  analyse the growth and the evolution of the  bar
instability in stellar-gaseous disks embedded in a DM halo evolving in a 
fully consistent cosmological scenario. 
Our model cannot be viewed as a general galaxy evolution model, since the
gradual formation and growth of the stellar disk has not been taken into
account. 
 Thus our work cannot be compared with the more recent papers by  
\citet{Aba03}, \citet{Gov04}, \citet{Spri04}, \citet{SLres} where
the formation of a realistic disk galaxy within the hierarchical scenario of
structure formation in {$\Lambda$ }-dominated cosmologies has been followed
 self-consistently. 
However our approach allows us to vary parameters like the
disk-to-halo mass ratio and the gas fraction inside the disk, and to 
analyse the growth of the bar instability and its dependence on such parameters 
 in a self-consistent cosmological framework for the first time. \\
Following the pioneering study by  \citet{ber98}, we compare the results of
such new set of simulations with those of a previous set \citep{Cu06}
including a
pure stellar disk in the same scenario
and with the same initial conditions.\\
The plan of the paper is the following: in Section 2, we  summarise
our recipe for the initial $disk+halo$ system fully described in Paper I.
In Section 3 we present the cosmological simulations and in Section 4 we point
out our  results.
 Section 5 is devoted to our discussion and conclusions. 

 \section{ Method}

We embed a gaseous and stellar disk inside a  cosmological halo selected in a
suitable slice of Universe and follow its  evolution inside a cosmological
framework: a $\Lambda$CDM  model with
$\Omega_{m}=$0.3, $\Omega_{\Lambda}=$0.7, $\sigma_8=$0.9, 
$h=$0.7, where $\Omega_{m}$ is
the total matter of the Universe, $\Omega_{\Lambda}$ the
cosmological constant, $\sigma_8$  the normalisation of the power spectrum,  
and $h$ the value of the Hubble constant in units of 100$h^{-1}$
km\,s$^{-1}$Mpc$^{-1}$. 

A detailed description of our method to produce the cosmological
scenario where the disk is evolved has been  given in \citet{Cu06}.
Here we present a short  overview of our  recipe.
Moreover, \citet{Cu06} showed that the numerical resolution does not
impair  their main result: in pure stellar disks, long living bars are a 
'natural' outcome of the cosmological scenario.  

\subsection{The DM halo}

To select the DM halo, we perform a low-resolution 
simulation of a ``concordance'' $\Lambda$CDM cosmological model. 
The initial redshift is 20.

From this simulation
we identify  the DM halos  at z=0  in the mass 
 \footnote{In the following, we  refer to the mass as
the virial mass i.e. that enclosed in a sphere with overdensity
$ \delta = \rho /\rho _{crit}=178\cdot \Omega _{m}^{0.44}$ 
\citep{Nav00}.} range
0.5- 5\( \cdot  \)10\( ^{11} \)h\( ^{-1} \) M\( _{\odot },\) with a standard  
friends-of-friends algorithm.\\
 We select one suitable DM halo with a mass
 M\( \sim  \)10\( ^{11} \)\( h^{-1} \) M\( _{\odot } \) (at z=0). We 
resample it with the multi-mass technique 
described in \citet{Kly01}. The particles of the DM halo, and those
belonging to a sphere with a radius $4 h^{-1}$ Mpc,  are followed
to their Lagrangian position and re-sampled to an equivalent resolution of 
1024 \(^{3} \) particles. 
The total number of DM particles in the high resolution region
is $1216512$ which corresponds to a DM  mass resolution of
1.21\,$10^{6} h^{-1}\,M_{\odot }$. 
The high-resolution DM halo is followed to the redshift z=0.

After selecting the halo and resampling at the higher resolution
the corresponding Lagrangian region, we run  the DM simulation, to extract 
the halo properties in absence of any embedded stellar disk.
The mass of our  halo at z=0,
\(1.03\cdot 10^{11}h^{-1} \) M\( _{\odot } \), corresponds to a
 radius,  \( R_{vir} = 94.7h^{-1}\) Kpc, which entails
84720 halo particles.
The nearest DM halo
\footnote{Halos have been identified
using the friends of friends algorithm with a linking length $l = 0.15$, i.e.
the mean
interparticle distances, and with more than 8 particles}
 more massive than $10^{10} h^{-1} M_\odot$ is 
$\sim 1900 h^{-1}$ Kpc away from the centre of our halo;
 the less massive one,  having mass of $4.6\cdot 10^7 h^{-1} M_{\odot}$,
is $\sim 215 h^{-1}$ Kpc away. Moreover, the behaviour of
the density contrast, $\delta$, is
monotonically decreasing with the radius, and  $\delta$ falls below the
unity value at 
$\sim ~550h^{-1}, ~450h^{-1}, ~350 h^{-1}$ physical Kpc away from the centre of our
halo at $z=0$, $z=1$, and $z=2$ respectively.
Therefore, we conclude that the selected halo is  living in an 
underdense environment.
From its accretion history (see Fig. 1 in \citet{Cu06}) we conclude that our halo undergoes no 
significant merger during the time it hosts our disk, nor immediately before.
The halo density profile  is well--fitted by a Navarro, Frenk and White
(NFW) form (\citet{Nav96}; \citet{Nav97})
at $z \leq 2$. The concentration, $C_{vir}$, equal to
\footnote{ We note, however, that $C_{NFW}$
is defined  against $R_{200}$, the radius enclosing a sphere with overdensity equal to
200 times the critical density of the Universe, and not against $R_{vir}$ as here;
therefore, in our cosmological model, it is always $C_{NFW}<C_{vir}$.
At $z=0$ our halo has $C_{NFW} \sim 14.$} 
$R_{vir}/R_s$, takes an high value, 18.1,  confirming that this halo does ``form'' 
at  quite high redshift  ( e.g. \citet{Wec02} for a discussion about
the link between  concentration and  assembly history of the halo).
The dimensionless spin parameter of the halo
is 0.04 at z=2, near to the average one for our cosmological
model \citep{Mal02}.\\

\subsection{The baryonic disk}

The spatial distribution of  particles  follows the
exponential surface density law: \( \rho =\rho _{0}\exp -(r/r_{0}) \)
where \( r_{0} \) is the disk scale length, \( r_{0}=4h^{-1} \) Kpc, and \( \rho _{0} \)
is the surface central density. The disk is truncated at five scale lengths with
a radius:
\( R_{disk}=20h^{-1} \) Kpc. To obtain each disk particle's position according
to the assumed density distribution, we used the rejection method \citep{Pre86}.
We used 56000 star particles and 56000 gaseous particles to describe our disk.
The (Plummer equivalent) softening length,  the same for DM, gas, and star particles,
is $ 0.5 h^{-1}\,Kpc$ in comoving coordinates.
The stellar radial velocity dispersion, ${\sigma}_R$,  is assigned through a 
Toomre parameter Q. Q  is
initially constant at all disk radii and it is  defined as 
${Q}= {{ {\sigma}_R \,\kappa } \over{ 3.36 \,G \,\Sigma }}$, where 
 $\kappa$ is the epicyclic
frequency, and ${\Sigma}$  the surface density of the disk. 
We plug a value of Q: 1.5, which corresponds to a {\it  warm}
disk.
The gas particles share the same velocity field as the stars.  The constraint
$Q_g > 1$ where $ Q_g = {{v_{sound}\, \kappa}\over{\pi G \Sigma_g}}$ , needed for the gaseous disk stability \citep{friebe93} is always verified.

 We embed the disk in the high resolution cosmological  simulation, at 
 redshift 2, in a plane perpendicular to the  angular momentum vector of
 the halo and in gravitational equilibrium with the potential.  Its centre of 
mass corresponds to the the minimum potential well of the DM halo.
The  initial redshift corresponds to 10.24 Gyr down to $z=0$ in our chosen cosmology.

\section{Simulations}

We performed eight cosmological simulations of a disk+halo system.
We also performed one  simulation in an isolated framework, using a
NFW halo,  to disentangle the effects of
the  dynamical state of the halo, and in particular of the evolutionary 
framework, on the growth of the bar instability (\S\,\ref{morf}).
We exploited the public parallel N-body treecode GADGET-2 \citep{Spri05}.

The simulations run on the SP4 and CLX computers located at 
the CINECA computing centre (BO, Italy; grant inato003) and on OATo Beowulf-class cluster
of 32 Linux-based PC (16 processor AMD K7, clock 700 MHz, and 16 Pentium 4, clock 1700
MHz) at the Osservatorio Astronomico di Torino.
\begin{table*}
\begin{center}
\caption{ Simulations: initial values}
\label{cosmsimtable}
\begin{tabular}{c c c c c c c c}
\hline\hline
N  & M$_{disk}$ & gas fraction & M$_{DM}$ & R$_{DM}$ & $\alpha \, r_m$ &${v_m}\over{{(\alpha G M_{disk})}^{1/2}}$ & halo\\  
\hline
 c1 &  0.33 & 0.1 & 0.64 & 1.94 & 1. & 1.08 & cosmological \\
 c2 &  0.33 &  0.2 & 0.64 & 1.94 & 1. & 1.08 & cosmological \\
 c3 &  0.33 & 0.4  & 0.64 & 1.94 & 1. & 1.08 & cosmological \\
 c4 &  0.1 & 0.1 & 0.64 & 6.4 & 0.9 & 1.68 & cosmological \\
 c5 &  0.1 & 0.2 & 0.64 & 6.4  & 0.9 & 1.68 & cosmological \\
 c6 &  0.1 & 0.4 & 0.64 & 6.4  &  0.9 & 1.68 & cosmological \\
 c7 &  0.1 & 0.5 &  0.64 & 6.4  &  0.9 & 1.68 & cosmological \\
 c8 &  0.1 & 0.6 &  0.64 & 6.4  &  0.9 & 1.68 & cosmological \\
 i1 &  0.33 & $ 0.2 $ & 0.95 & 2.87 & 0.85 & 1.5 & NFW \\
\hline
\end{tabular}
\end{center}
I   col: simulation number and simulation type (c: cosmological simulation,
i: isolated simulation) \\
II col: mass of the disk in code units (i.e. $5.9\times 10^{10}\, M\odot$)\\
III  col: fraction of gas, i.e. gas--to--disk mass ratio\\
IV  col: DM mass inside the disk radius in code units\\
V   col: halo--to--disk mass ratio inside the disk radius\\
VI and VII cols: \citet{Efs82} parameters: $\alpha={r_0}^{-1}$, $v_m$ is
the maximum rotational velocity, $r_m$ the corresponding radius\\
VIII col: type of halo used 
\end{table*}
\begin{table}
\begin{center}
\caption{ Simulations: final results}
\label{cosmsimtable_fin}
\begin{tabular}{c c c c c c c c }
\hline\hline
 N & M$_{DM}$ & R$_{DM}$ &  $\epsilon$ & Q$_b$  & a$_{max}$ &  bulge & bars\,in\,bars\\  
\hline
 c1 & 0.77 & 2.39 & 0.68 & 0.43 & 8.4 & y & n\\
 c2 & 0.78 & 2.4 & 0.1 & 0.01 & n & n & n\\
 c3 & 0.78 & 2.4 & 0.07 & 0 & n & n & n\\
 c4 & 0.73 & 7.43 & 0.58 & 0.32 & 5.8 & n & y\\
 c5 & 0.73 & 7.44 & 0.6 & 0.3 & 5.4 & n & n \\
 c6 & 0.73 & 7.56 & 0.46  & 0.23  &  5.6 & n & n\\
 c7 & 0.73 & 7.6  &  0.46  &   0.2 & 5.4  & n & n  \\
 c8 & 0.73 & 7.6  &  0.42  &   0.21     &   5.8  & n    & n  \\
 i1 & 1.   &  3.03 & n & n &  n & n & n \\
\hline
\end{tabular}
\end{center}
I   col: simulation number and simulation type\\
II  col: DM mass inside the disk radius in code units\\
III   col: halo--to--disk mass ratio  inside the disk radius\\
IV col: maximum ellipticity at z$=0$: strong
bar \citep{Ma01} require $ \epsilon \ge 0.4$\\
V col: bar strength  according to \citet{Comb81}; stronger bars
correspond to higher Q$_b$ values\\
VI col: major axis (physical Kpc) corresponding to the maximum bar strength\\
VII col: morphology of the inner region of the disk\\
VIII col: peculiar features inside the disk 
\end{table}
The main parameters and the initial properties of our set of  simulations
 are listed in Table  \ref{cosmsimtable}.
 
A global stability criteria for the stellar bar instability in a disk galaxy
is the one analysed in  \citet{Efs82}. In such a paper the parameters  $\alpha
 \, r_m$ and  $v_m\over{{(\alpha M G)}^{1/2}} $ (where  $v_m$ is the maximum
 value of the disk rotational curve,  $r_m$ the corresponding radius,
 ${\alpha} = {{r_0} ^{-1}}$ and $M$ is the disk mass) have been defined.
\citet{Efs82} stated the criterion  ${v_m\over{{(\alpha M G)}^{1/2}}}  \geq
  {1.1} $ over the range  $0.1 \leq {\alpha \, r_m} \leq 1.3$ for a disk
 being stable to bar formation. The values of these parameters
are reported in Table  \ref{cosmsimtable}.
Simulation i1  corresponds to the evolution of the
same disk as in simulation c2 but  we
embedded it in a NFW halo  
with the same virial mass and particle number as in our cosmological halo at z=0 (see
\citet{Cu06} for more details). Such a halo is
spherical, isotropic and in gravitational equilibrium.  The procedure used to
construct this halo is described by \citet{hern93}.

All the results  in Table \ref{cosmsimtable} 
can be compared with those of
the homologous simulations performed with a  pure stellar disk,
presented in \citet{Cu06}. 
We  repeated, indeed,  one of the simulation of \citet{Cu06} with
the GADGET-2 code, since the time--stepping criterion we used is not
present any more in the current version of the code. A comparison with
the previous run does not show any significative change.
In such a paper we also performed several
 numerical tests on the stability of the results.
Here we summarise our findings:
\begin{itemize}
\item  a simulation performed with a mass resolution
 increased by a factor $8$, i.e. with a 
  force resolution increased by a factor $2$ (for a disk--to--halo mass ratio 
  0.1; see section A.2 in Curir et al. 2006), gives rise to a long living bar
  (10 Gyr old) as in the case of the lower mass and force resolution;
  note that in this test the mass and force resolution are respectively
  $M_{DM}^{part} \approx 2 \cdot 10^5 h^{-1} M_\odot$ and $\epsilon
  \approx 250 h^{-1}$ pc, thus comparable to the resolution of recent
  works on galaxy formation simulations, e.g. \citep{SLres,Stinson,Fabiores};
\item embedding adiabatically a classically stable stellar disk in a 
isolated DM halo
  having a NFW profile, while keeping the mass and force resolution
  constant, brings to {\it no} bar instability, thus ruling out the
  possibility that the embedding procedure is the main driver of the
  bar instability (see Section 5 in Curir et al. 2006); 
\item varying the {\it stellar} softening, at fixed mass resolution,
  from 0.36h$^{-1}$Kpc to 0.65h$^{-1}$Kpc does not result in any
  significant changes as far as the bar instability is concerned (see section
  A.3 in Curir et al. 2006);
\item the same holds when the stellar mass resolution is reduced by a factor
  of 1/6 and increased up to 10 times the value adopted in our set of 
   simulations (see A.2. in  Curir et al. 2006).
\end{itemize}
These results   guarantee that, in the dynamical range
currently  reached with tree--code N--body simulations, no mayor
numerical shortcoming is expected, at
least under the point of view of the gravitational evolution.
Moreover, as a final test,  we re-run simulation c2  using two times more 
gaseous particles, 112000 instead of 56000. We find that
the final bar dissolution and the gas inflow are the same as in 
the previous case.

\section{Results}

In this section we  present  the final ( i.e. at z=0)
isodensity contours for  stars and gas of simulations in Table
\ref{cosmsimtable}.  All such figures have been built up with the same
box-size, number of levels, and density contrast (see caption of Fig.
\ref{dens1}).\\
 We define  as a measure of the bar strength the value of 
the  ellipticity, $\epsilon=1-b/a$ (Table \ref{cosmsimtable_fin}); 
a strong bar corresponds to  $\epsilon \geq 0.4$.

A more dynamical measure of the bar strength at radius R
has been defined by \citet{Comb81} by using  the parameter:
 $Q_t(R)= {F_T^{max}(R)\over{<F_R(R)>}}$
where $F_T^{max}(R)=[{\partial \Phi(R,\theta)}/{\partial \theta}]_{max}$ is 
 the maximum amplitude of tangential force  at a given radius R and 
 $<F_R(R)>=R({\partial \Phi_0}/{\partial R})$ is 
the mean  axisymmetric radial force derived from the $m=0$
component of the gravitational potential  at the same radius.
 The maximum value of  $Q_t(R)$ provides a measure of the bar strength $Q_b$ for  the whole galaxy.
We evaluated the components of the gravitational force  directly  from the
  masses and the radii of the system produced by the simulations.  We then
  represented the  matrix  $Q_t(i,j) = {F_T(i,j)\over{<F_R(i,j)>}} $  on a
suitable two dimensional grid  in analogy with  the method outlined by
  \citet{BuBlo01}. Naming   $Q_{bk}$ the maximum value of $Q_t(i,j)$ in
  quadrant k of the
  map, according with \citet{BuBlo01}, we define: $Q_b  = {{\sum_{k=1}^4 Q_{bk}}\over{4}} $

We listed the final values  of the parameter $Q_b$
in Table \ref{cosmsimtable_fin}.
The   values of the  bar  strength derived with both methods are consistent.

In all our cosmological simulations, with the exception of 
simulations c2  and c3,
a stellar bar is still living at z=0.
\subsection{Star and gas morphologies}\label{morf}

\begin{figure*}
\begin{center}
\includegraphics[width=17cm]{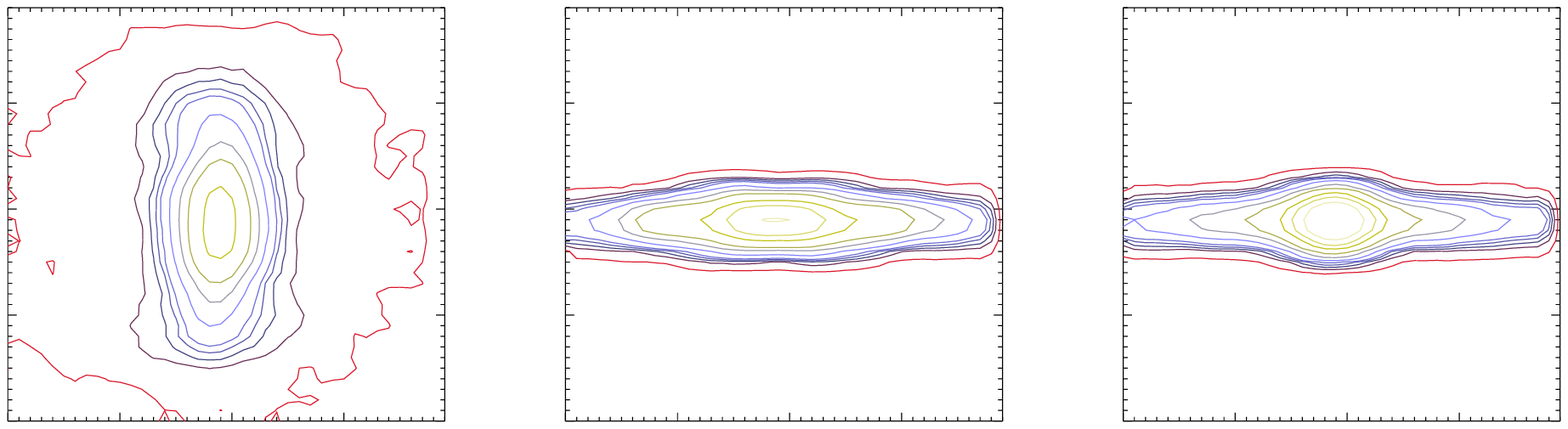}
\includegraphics[width=17cm]{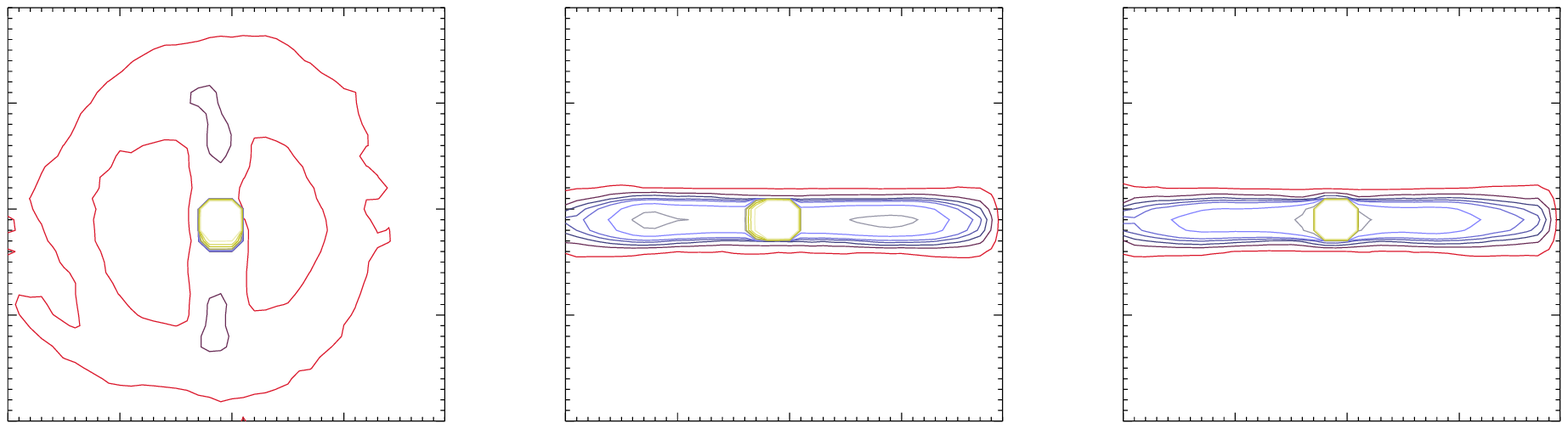}
\end{center}
\caption{ Face-on, edge-on and side-on isodensity contours (from left to right)
of simulation c1 at z=0 (see text); top panel shows the stellar component,
 bottom panel the
gaseous component. Spatial resolution  is always 
0.5$h^{-1}$ {\it physical \,} Kpc and the
box size is 40 times the spatial resolution.
Contours are computed at 11 fixed levels ranging from  
$2\times 10^{-4}$  to $0.015$  in term of density  fraction of stars  
(gas)  within the  spatial resolution to the total star (gas) density in the map.}
\label{dens1}
\end{figure*}

\begin{figure}
\begin{center}
\includegraphics[width=8cm]{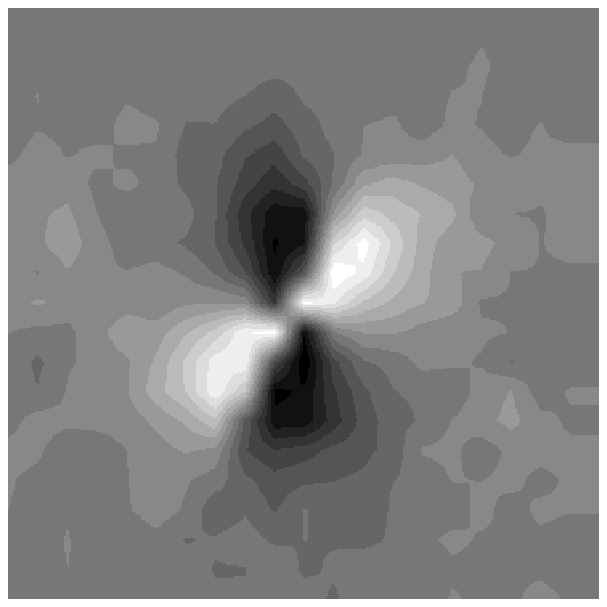}
\includegraphics[width=8cm]{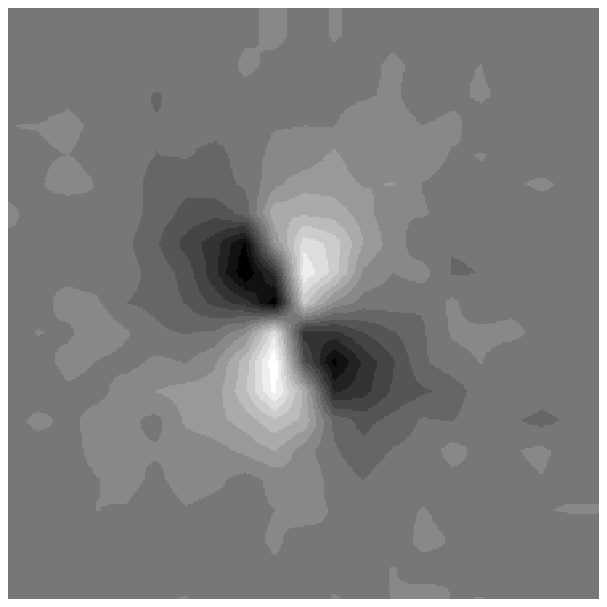}
\end{center}
\caption{ Maps of the ratios of the axisymmetric radial and tangential forces
 in the disk plane at z=0 for simulation c1  (top panel) and the analogous case without
  gas (bottom panel).
Box size and spatial resolution are the same as in Fig.1.}
\label{ratiomap}
\end{figure}

The strongest and the longest bar arises from simulation c1 (Fig. \ref{dens1}).  
In this case,
the ``geometrical''  strength of
such a bar,  namely the ellipticity, $1-{b/a}$, evaluated through  
the isodensity plots, is 
higher (0.68) than the one measured  in the homologous case 
(0.52, \citet{Cu06}). The strength measured through the $Q_b$ parameter is also  higher 
than the corresponding one in the non-dissipative homologous case as shown in 
Fig. \ref{ratiomap} which compares the maps of
$ Q_t(i,j) $ final values of the stellar bars.
The enhancement of the stellar bar strength  is due
to the gaseous bar which is superimposed, i.e coupled, to the stellar one along
all the evolution.
(see Fig. \ref{dens1}, bottom panel).
\begin{figure*}
\begin{center}
\includegraphics[width=6cm]{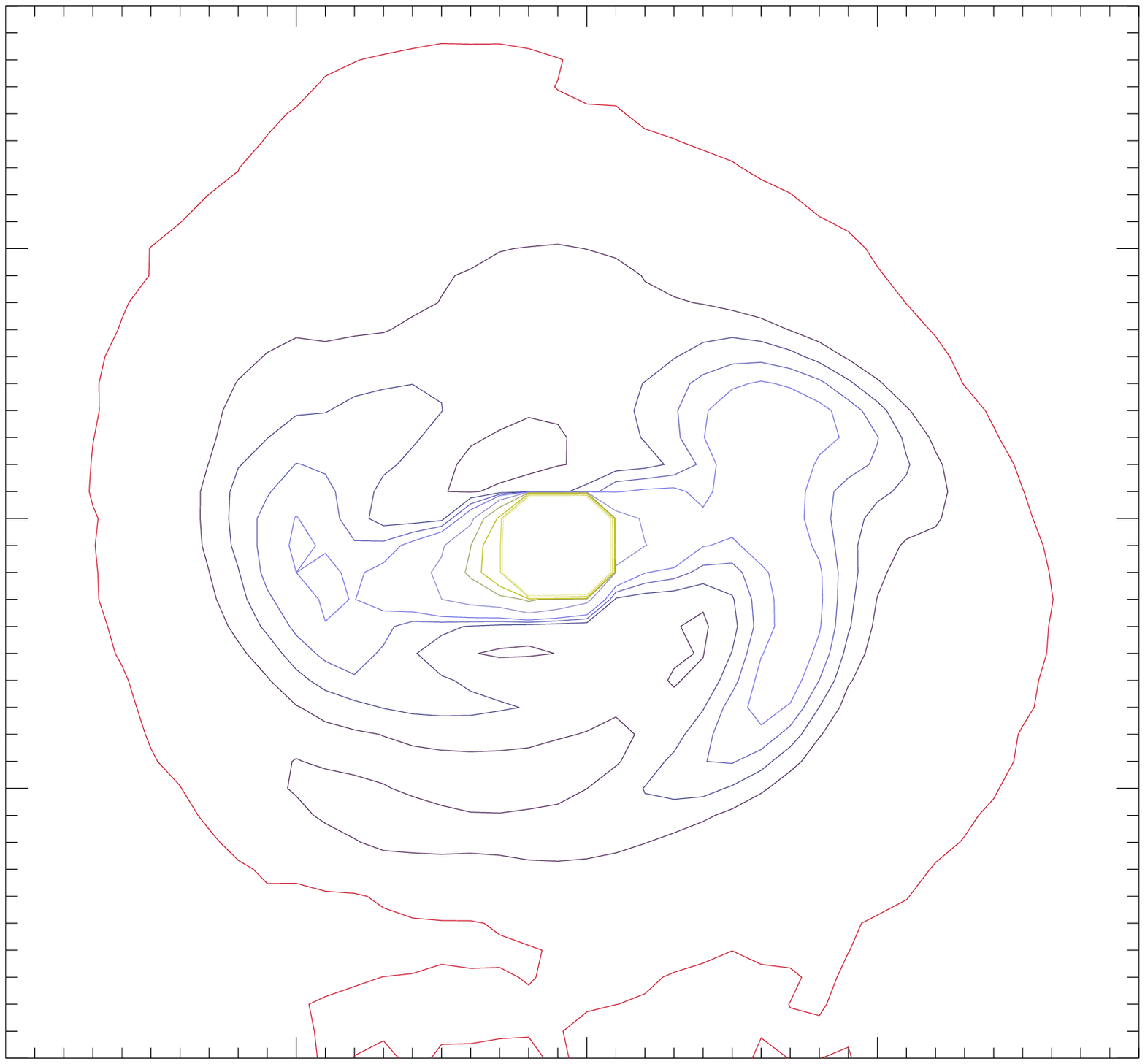}
\includegraphics[width=6cm]{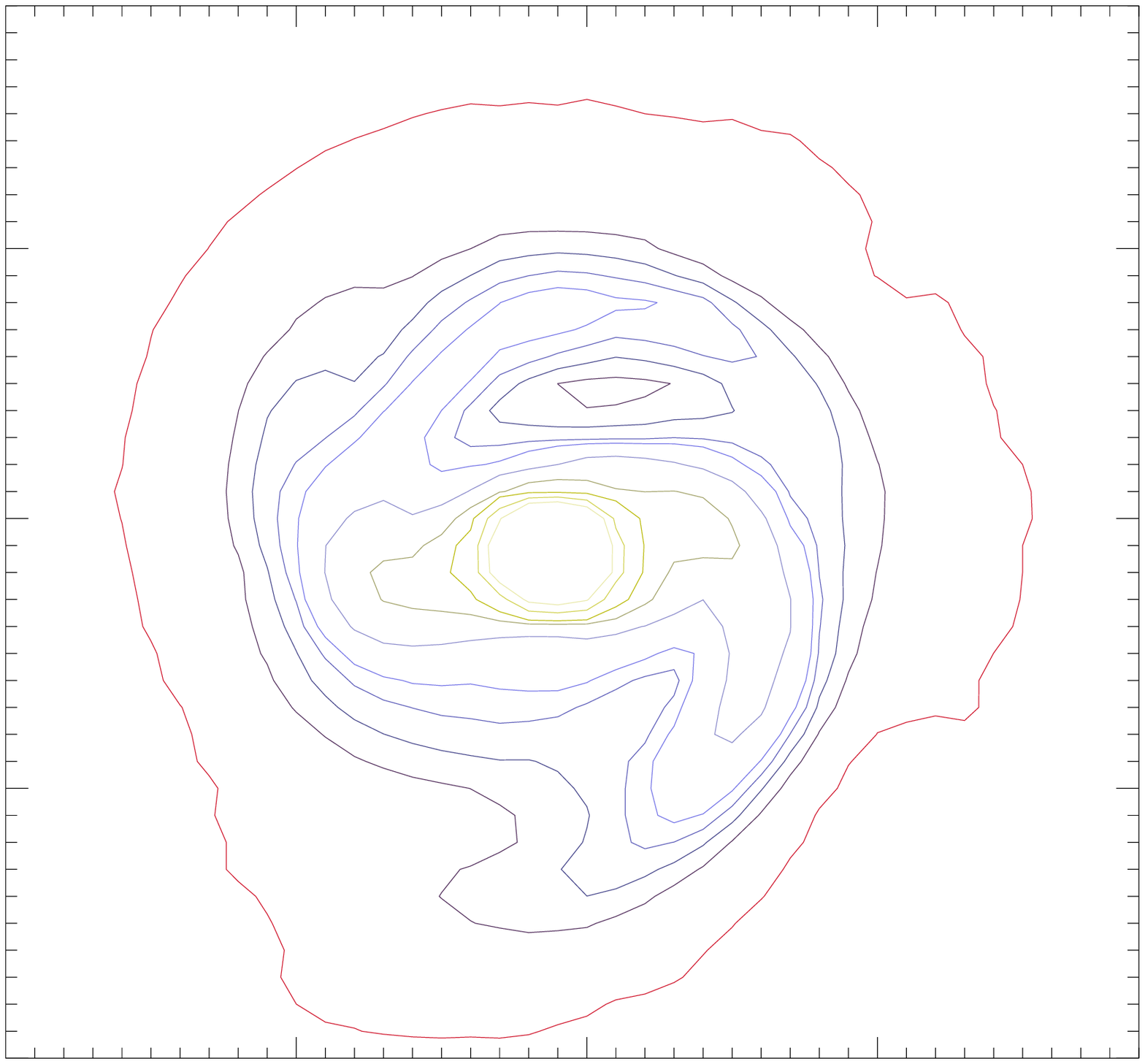}
\end{center}
\caption{ Gas isodensity contours,  x-y projection,
of simulation c1  (left panel) and c2 (right panel) at z=1.}
\label{gas_morfc1c2}
\end{figure*}
\begin{figure*}
\begin{center}
\includegraphics[width=17cm]{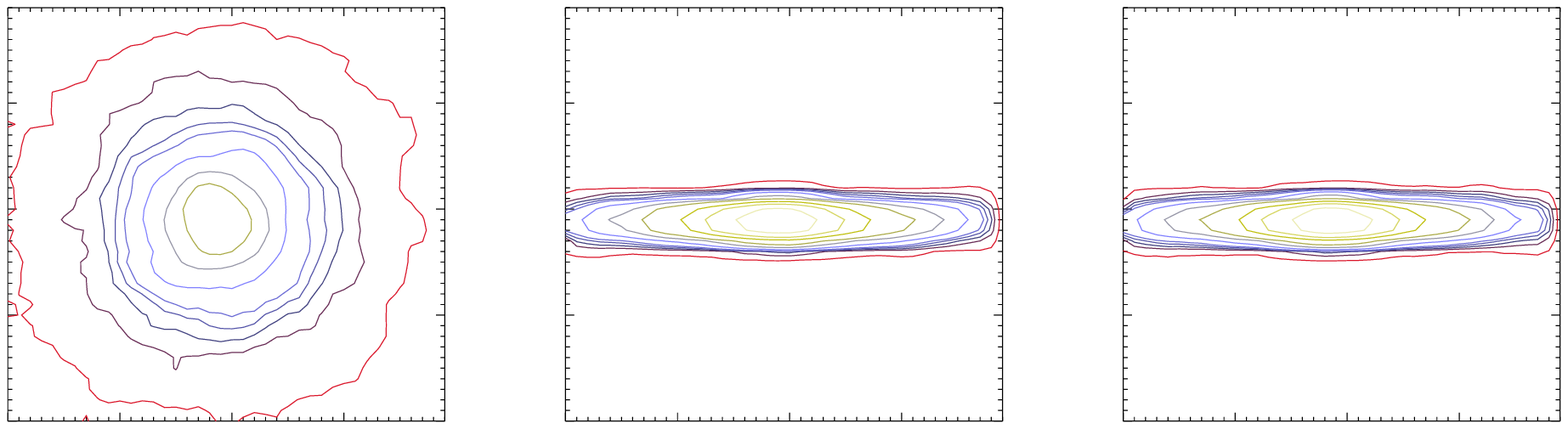}
\includegraphics[width=17cm]{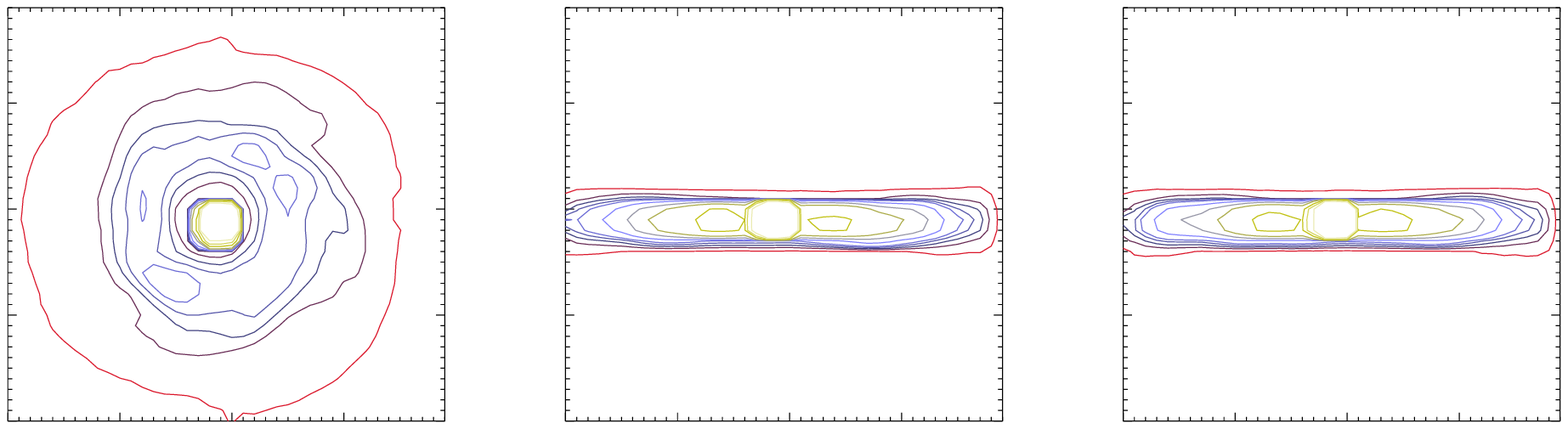}  
\end{center}
\caption{As in Fig. \ref{dens1} for  simulation c2.}
\label{dens2}
\end{figure*}
In the simulation c2, the gas develops a 'grand-design' spiral structure, more
evident than  in simulation c1 (Fig. \ref{gas_morfc1c2}).
Such an effect can contribute to destroy the stellar bar, in
agreement with recent findings by \citet{Bou05}. In the same simulation, c2, the
gas produces a barred  feature from the earlier stages,   lasting until
$z \approx 1.2$. After such a redshift, the gaseous central condensation 
  contributes to  the bar destruction, according to the claim of
\citet{ber98}. These authors show that the growth of a gas mass
concentration in the disk dissolves the regular orbits in the stellar bar. 
In such a simulation, the stellar
bar decreases its strength and disappears completely at $z \approx 0.15$,
whereas in the c3 case, which corresponds to a larger initial gas fraction, 
the bar disappears at $z \approx 0.6$.
Fig. \ref{dens2} shows the gas and star morphologies  of simulation c2 at z=0.
\begin{figure*}
\begin{center}
\includegraphics[width=17cm]{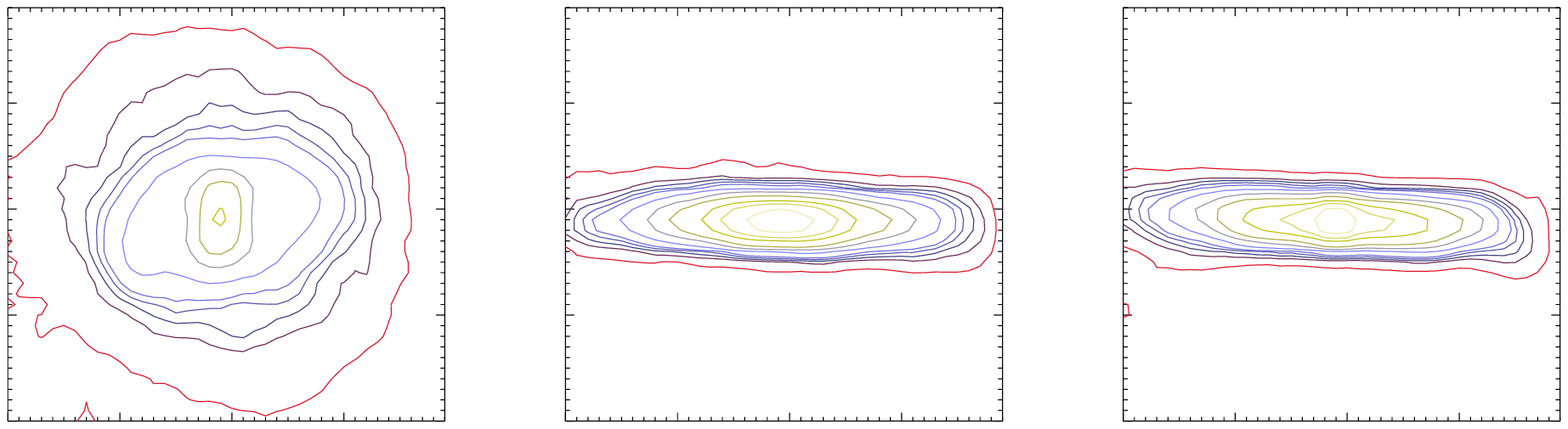}
\includegraphics[width=17cm]{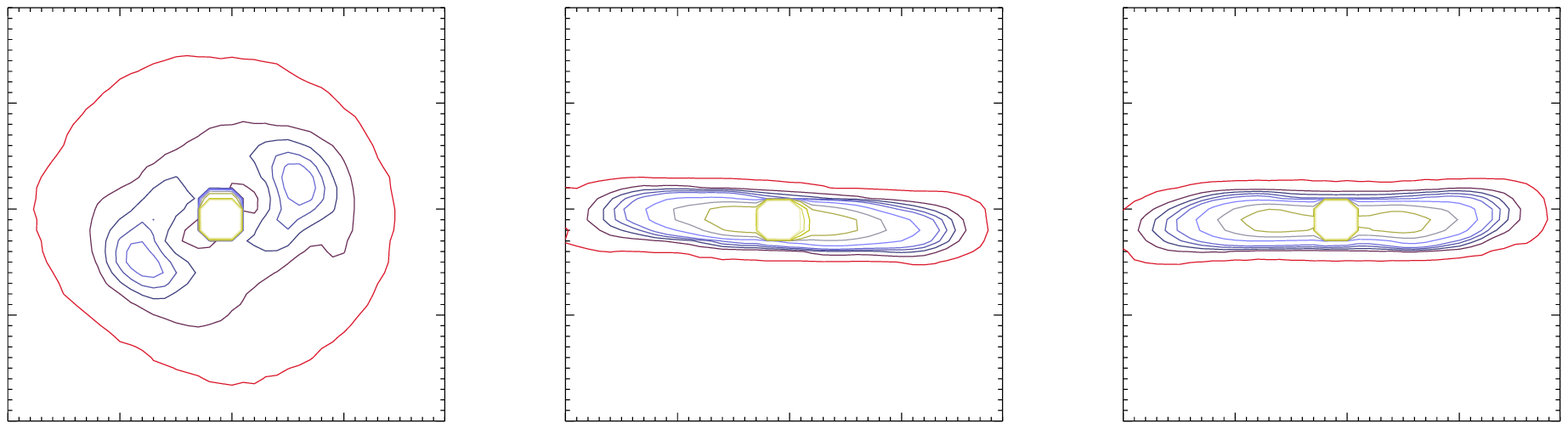}
\end{center}
\caption{ As in Fig. \ref{dens1} for  simulation c4.}
\label{dens3}
\end{figure*}
\begin{figure*}
\begin{center}
\includegraphics[width=17cm]{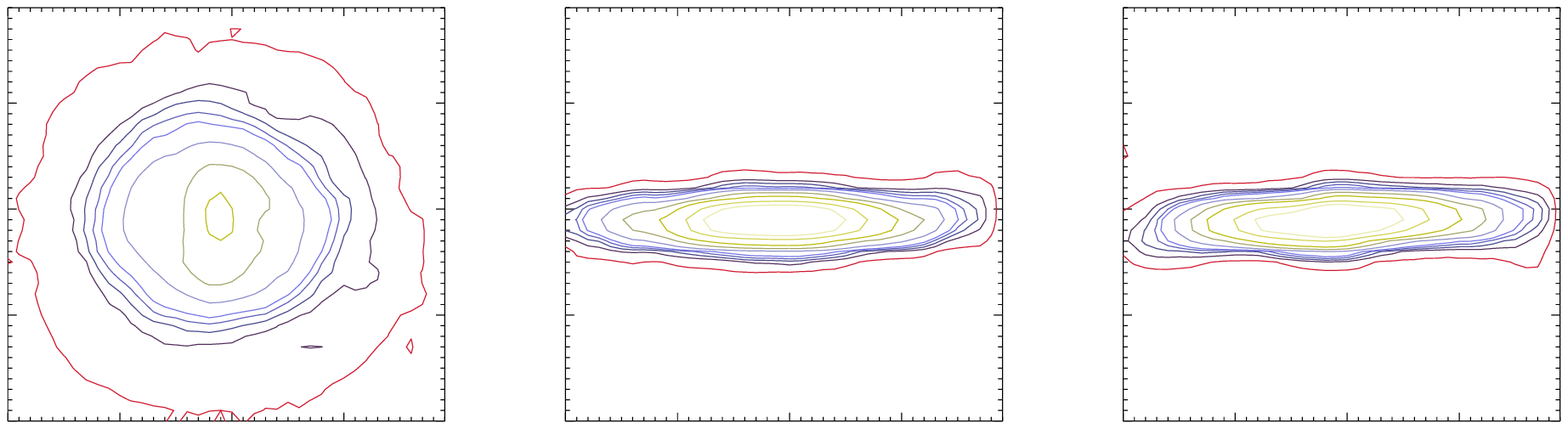}
\includegraphics[width=17cm]{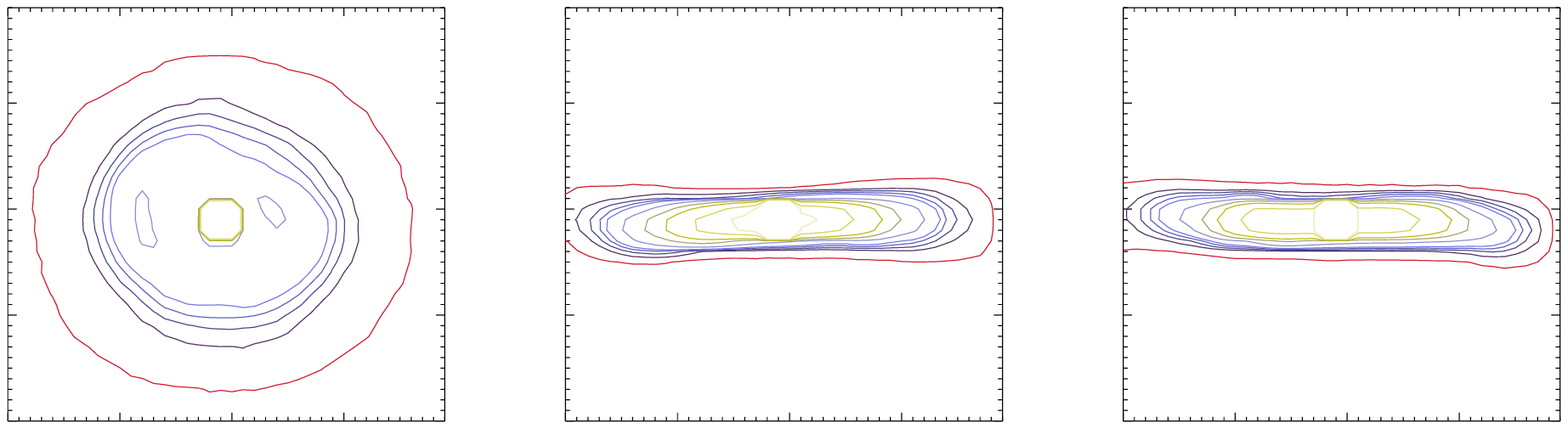}
\end{center}
\caption{As in Fig. \ref{dens1} for  simulation c8.}
\label{dens6}
\end{figure*}
\begin{figure*}
\begin{center}
\includegraphics[width=6cm]{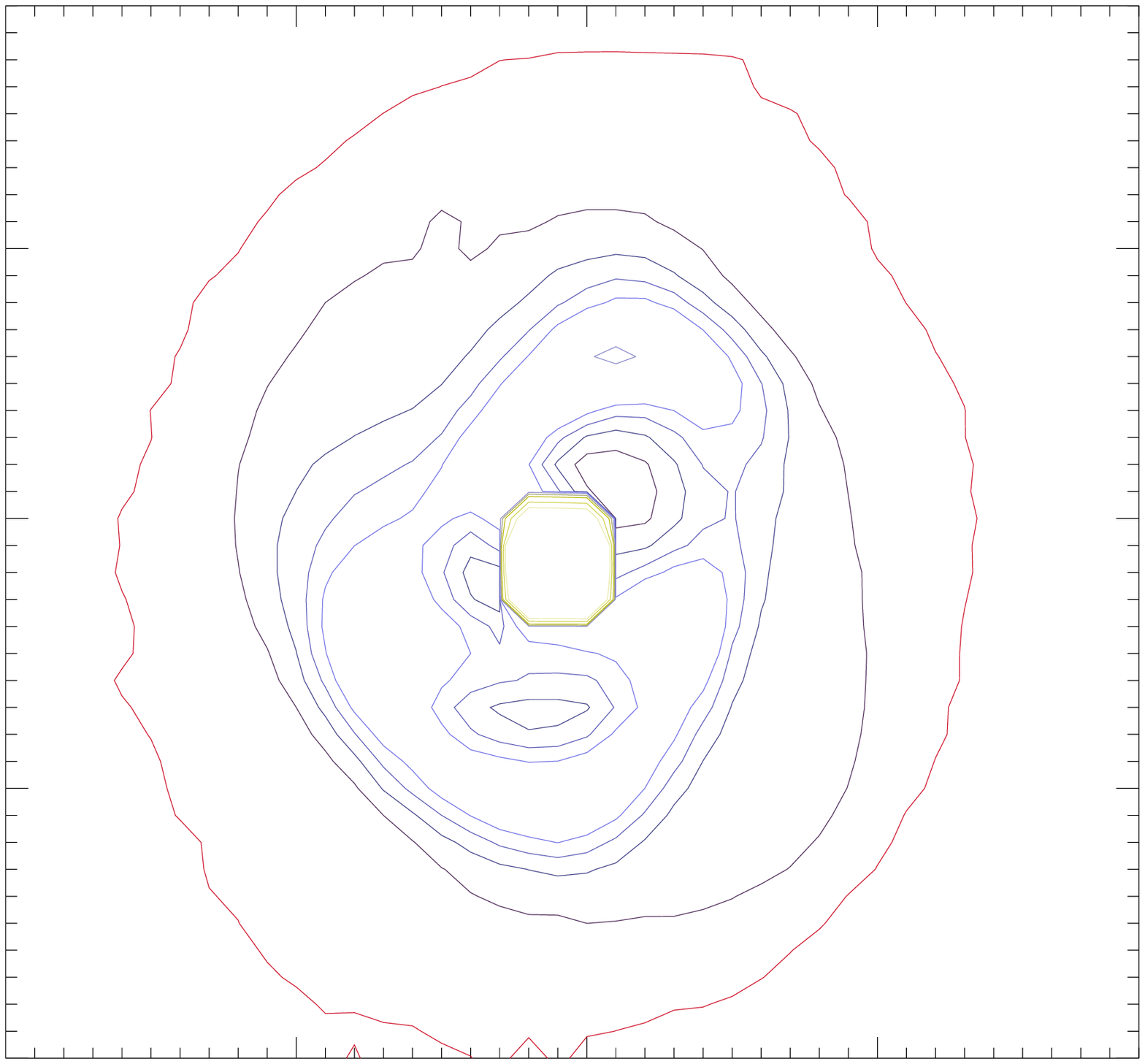}
\includegraphics[width=6cm]{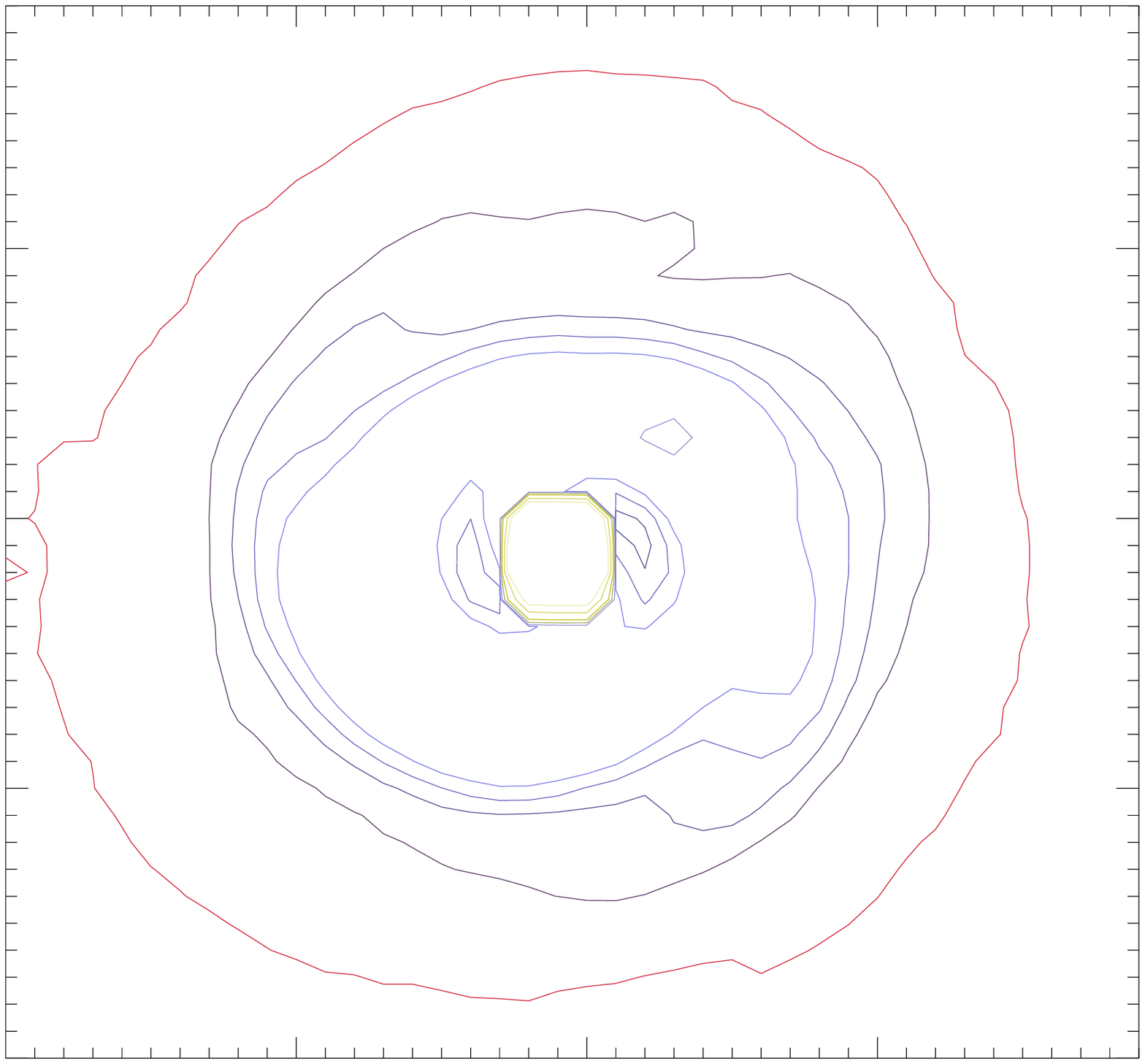}
\includegraphics[width=6cm]{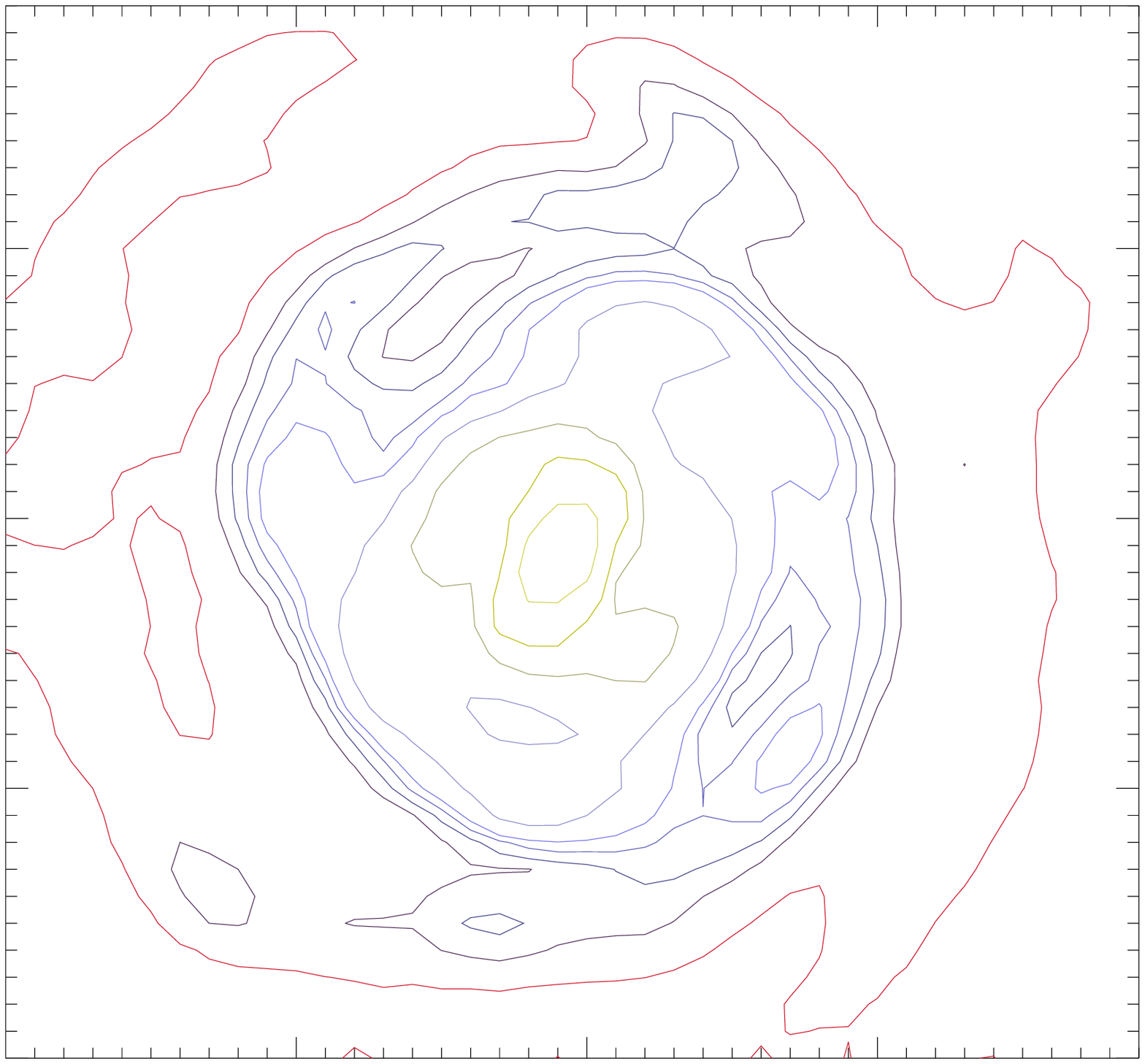}
\includegraphics[width=6cm]{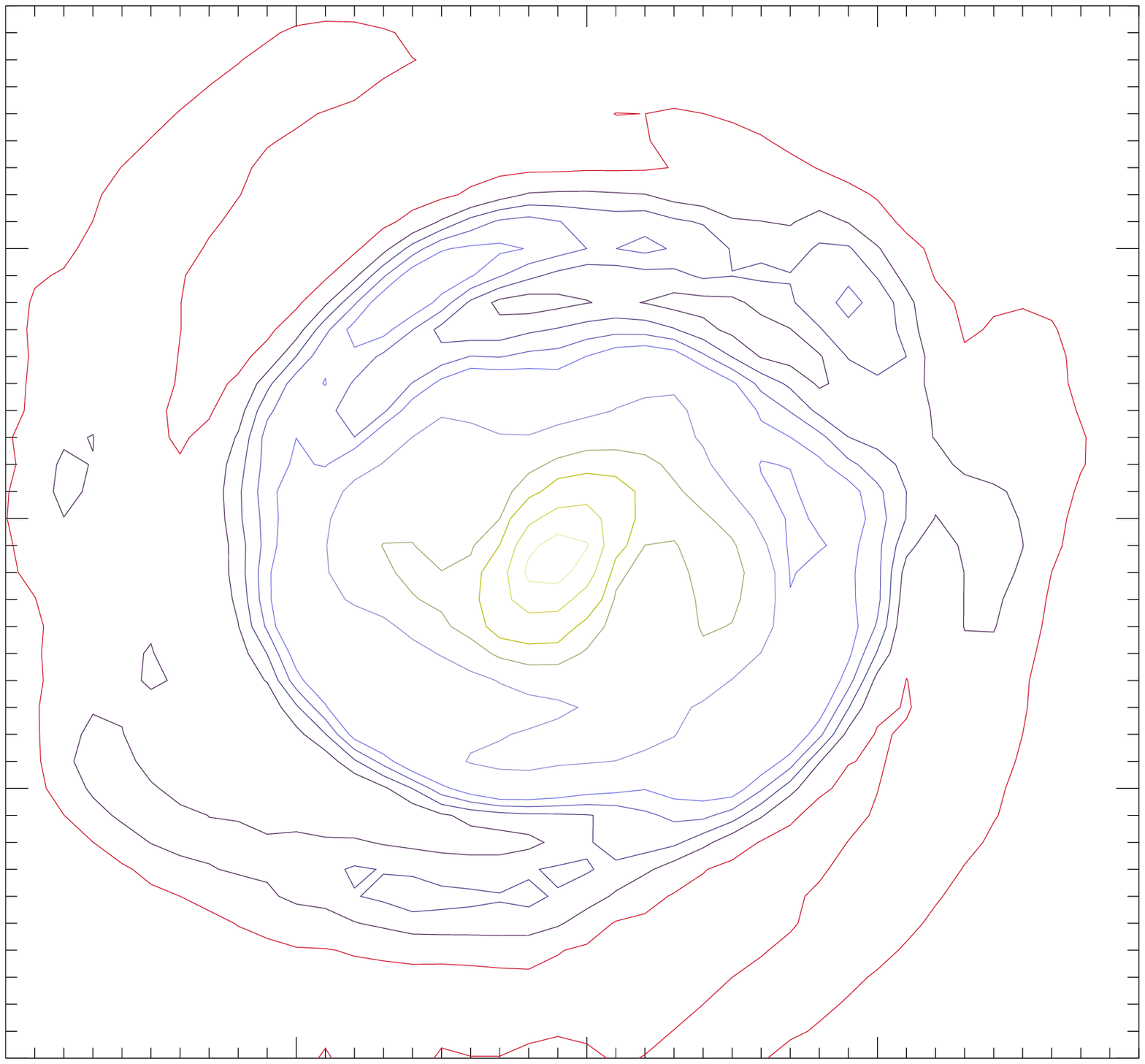}
\end{center}
\caption{ Gas isodensity contours (top panel),  x-y projection,
of simulation c4  (left panel) and c8 (right panel) at z=1;
bottom panels show that corresponding contours for stars.}
\label{comp}
\end{figure*}
\begin{figure}
\begin{center}
\includegraphics[width=8cm]{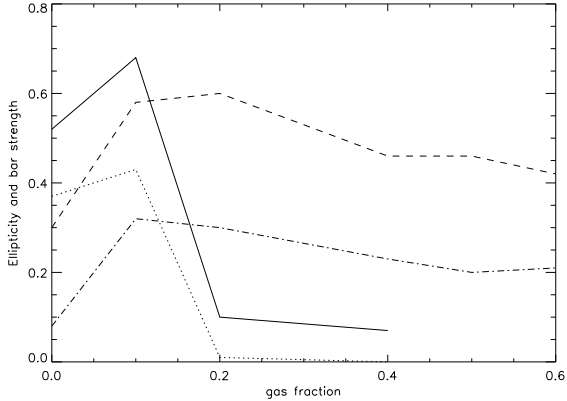}
\end{center}
\caption{ Behaviour of the bar strength and ellipticity at z$=0$ for our set of
cosmological simulations 
with increasing gas 
fraction. 
 $Q_b$ (dotted line) and ellipticity (full line) of our
more massive disks (i.e. disk--to--halo mass ratio
0.33); $Q_b$ (dot--dashed line) and ellipticity (dashed line) of our
less massive, DM dominated, disks (i.e. disk--to--halo mass ratio
0.1).
}
\label{strength}
\end{figure}
\begin{figure*}
\begin{center}
\includegraphics[width=17cm]{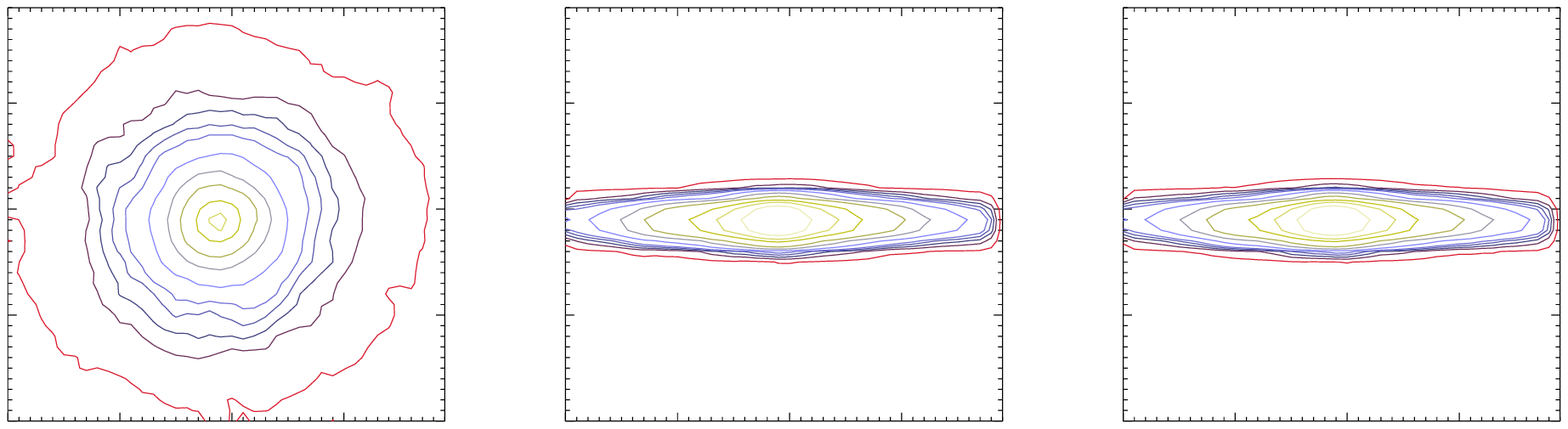}
\includegraphics[width=17cm]{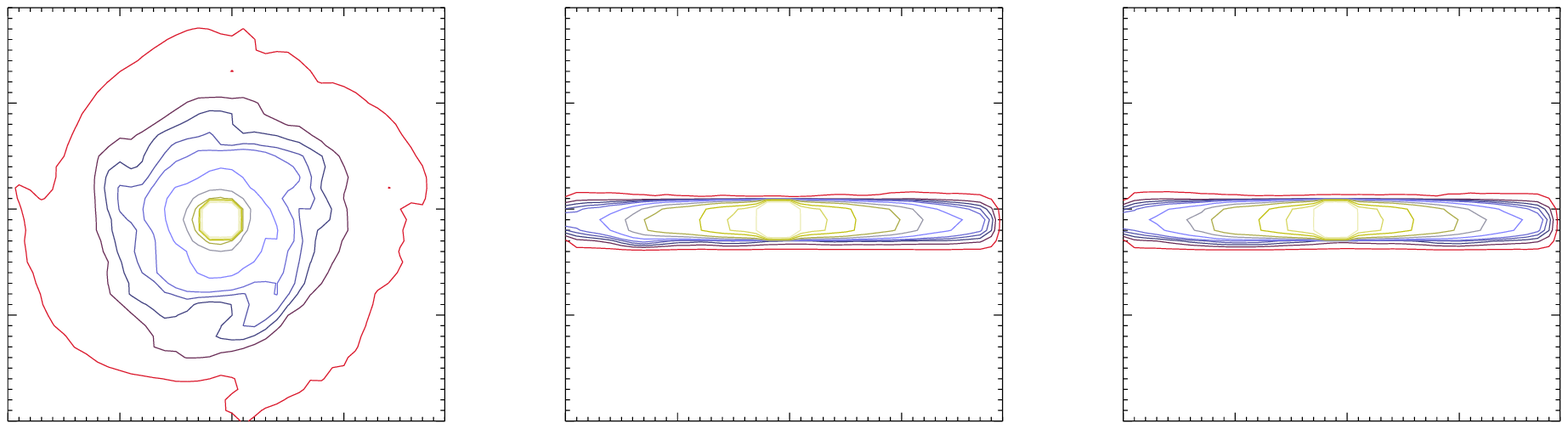}
\includegraphics[width=17cm]{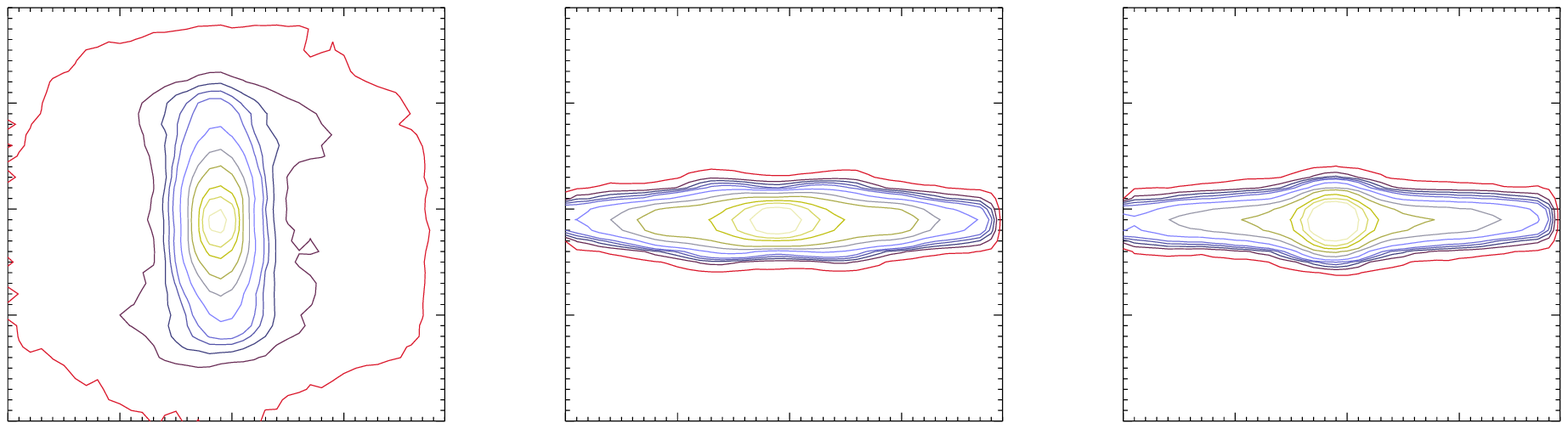}
\end{center}
\caption{Isodensity contours of  simulation i1 at 10.24 Gyr (top and central
panels) 
compared with the stellar isodensity contours of the same
case without gas (bottom panel, simulation c4 of \citet{Cu06}) at z=0.}
\label{NFW_morf}
\end{figure*}
Simulation c4  shows a  stronger stellar bar than in \citet{Cu06} (
ellipticity 0.58 from
Table \ref{cosmsimtable_fin} instead of 0.3  from their Table 3)
and  bar-in bar features  along the whole evolution until z=0  (see Fig.
\ref{dens3}). Simulations from c5 to c8, which correspond to increasing gas 
fractions, show 
the same but less enhanced features during the system evolution, so that at z=0
no bar-in-bar appear, and for simulation c8 (Fig. \ref{dens6}) the  final bar is the weaker one. 
In this set  of DM dominated disk simulations, the  action of 
the increasing gas mass
is  not enough to destroy the bar. We note  that, also in these cases, the   gaseous 
bar is coupled with that of the stars from the beginning, however simulations
with the larger gas fraction restore the gaseous disk earlier, so that 
gas and star distributions decouple from z=1 (Fig. \ref{comp}).

In Fig. \ref{strength} we show the behaviour of the ellipticity and of the
$Q_b$  as a function of the increasing mass fraction in disks of 
different masses.  It is remarkable the agreement we recover
between the trends: the dynamical evaluation of the bar strength and the geometrical
evaluation of it through ellipticities.

Fig.  \ref{NFW_morf} compares the gas and star morphologies  of
the non-cosmological simulation, i1, with those
of the homologous non-dissipative case (simulation i4 in Tables 2 and 3 of
\citet{Cu06}) after 10.24 Gyr.
In simulation i1 the bar forms approximately at the
same time as in the pure stellar case. However the presence of the 
gas destroys the bar after 7.5 Gyr instead of 8.3 Gyr as in simulation
c2. 
Thus the  cosmological framework fuels a longer living bar,
moreover, when the disk--to--halo mass ratio
is 0.33, in such a framework we find a critical value of the gas fraction, 
0.2, able to destroy the bar.

In the case of DM dominated disks, all the simulations here performed 
show a bar feature still present at z=0. This is enhanced for the lower  gas 
percentage,  0.1,  and slightly
weakened for the higher ones, compared with the corresponding no dissipative case
\citep{Cu06}, but the bar 
cannot be destroyed by the gas presence.

\subsection{Mass inflow   and circular velocities}
The gas, initially distributed over all the disk surface  with the same density
distribution as the stars, evolves by increasing its concentration
depending on the disk-to halo mass ratio and on its initial mass fraction.
Both such parameters affect
the life and the strength of the bar, as discussed above.

\begin{figure}
\begin{center}
\includegraphics[width=8cm]{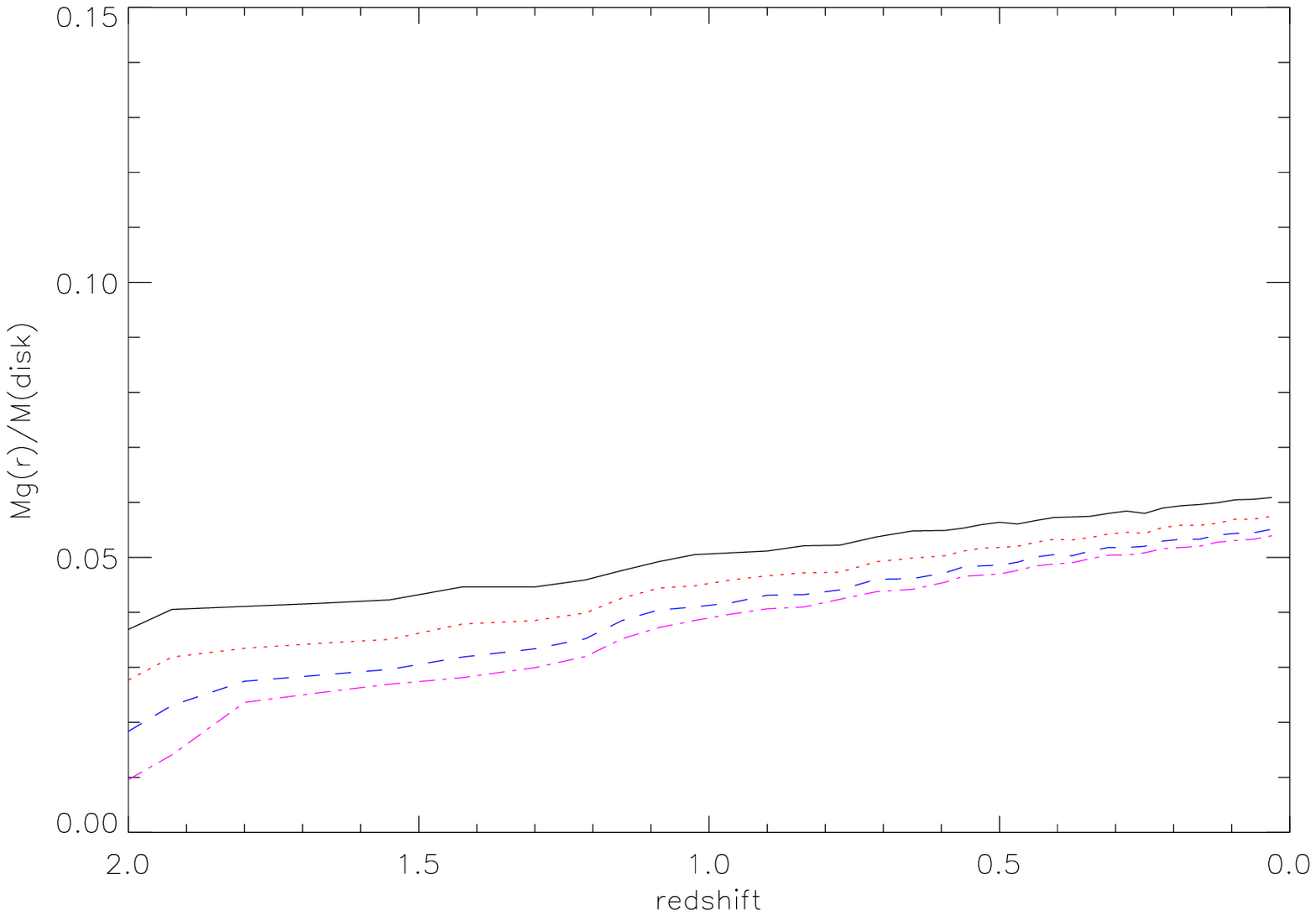}
\includegraphics[width=8cm]{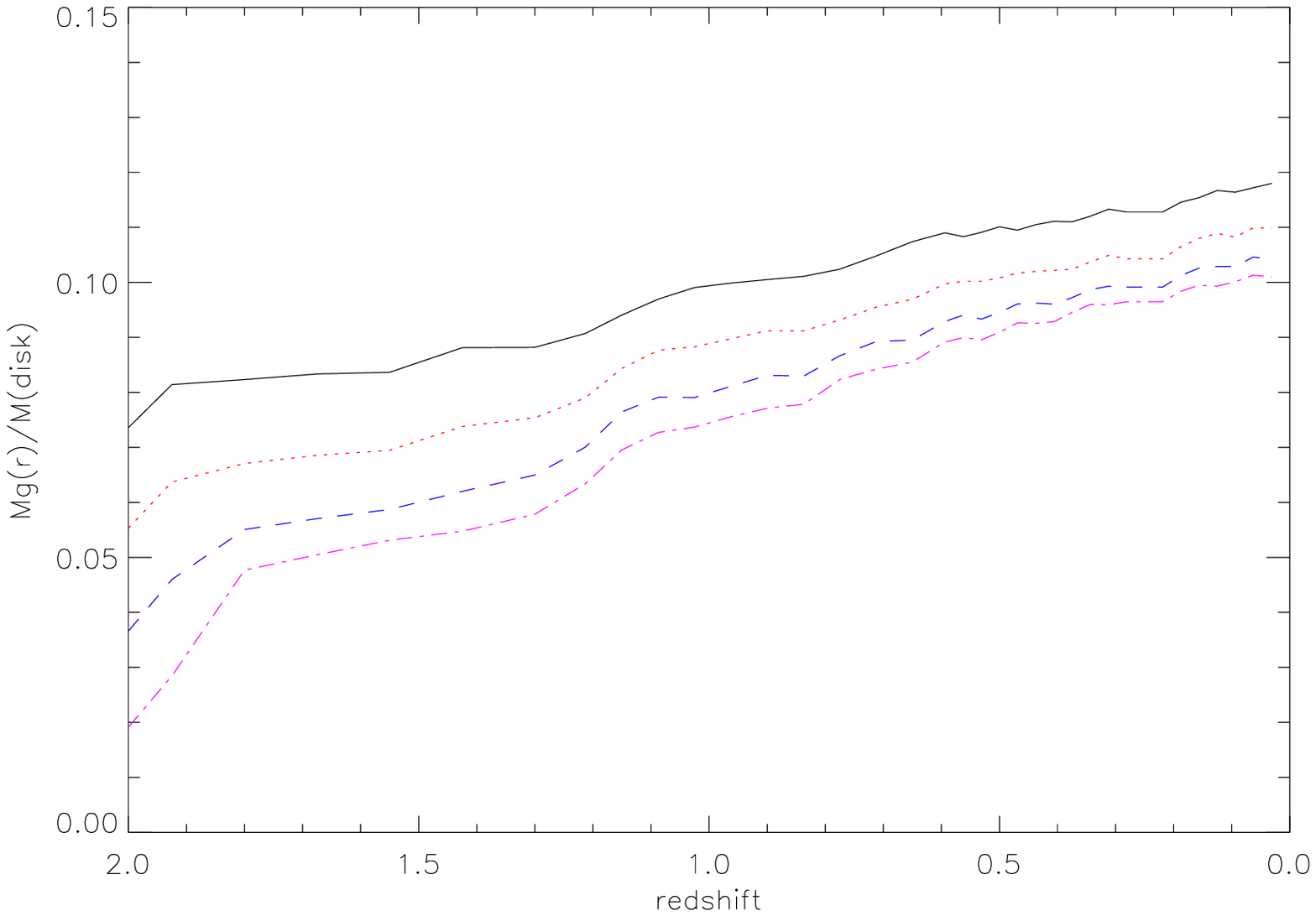}
\includegraphics[width=8cm]{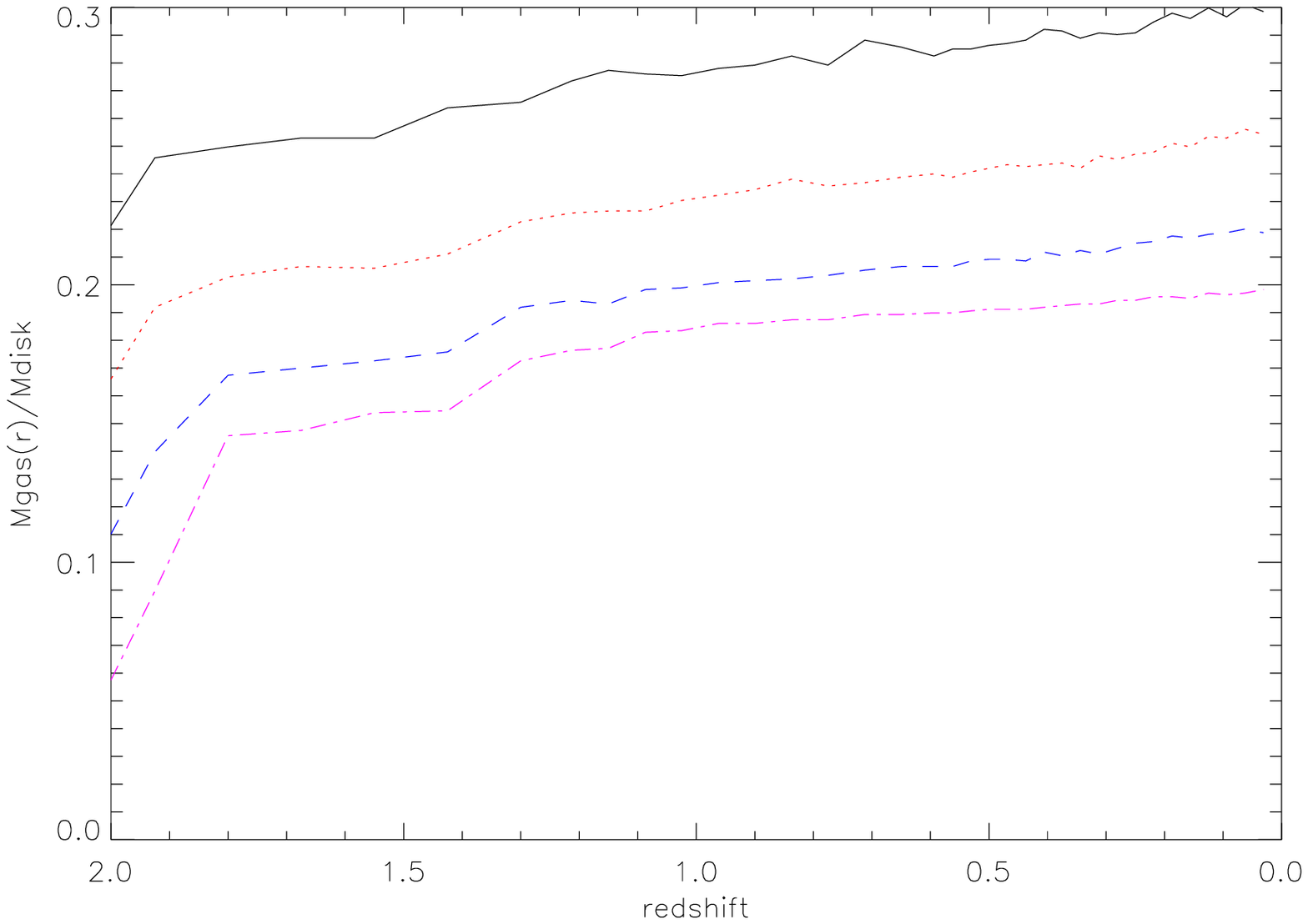}
\end{center}
\caption{ Gas--to--disk mass ratio at different disk radii: 2 Kpc 
(magenta dashed-dotted line), 3 Kpc
  (blue dashed line), 4 Kpc(red dotted line), and 5 kpc (black full line)  as function of
  redshift for   simulation c4 (top panel), c5 (middle
  panel), and c8 (bottom panel)}
\label{gas_m1}
\end{figure}

\begin{figure}
\begin{center}
\includegraphics[width=8cm]{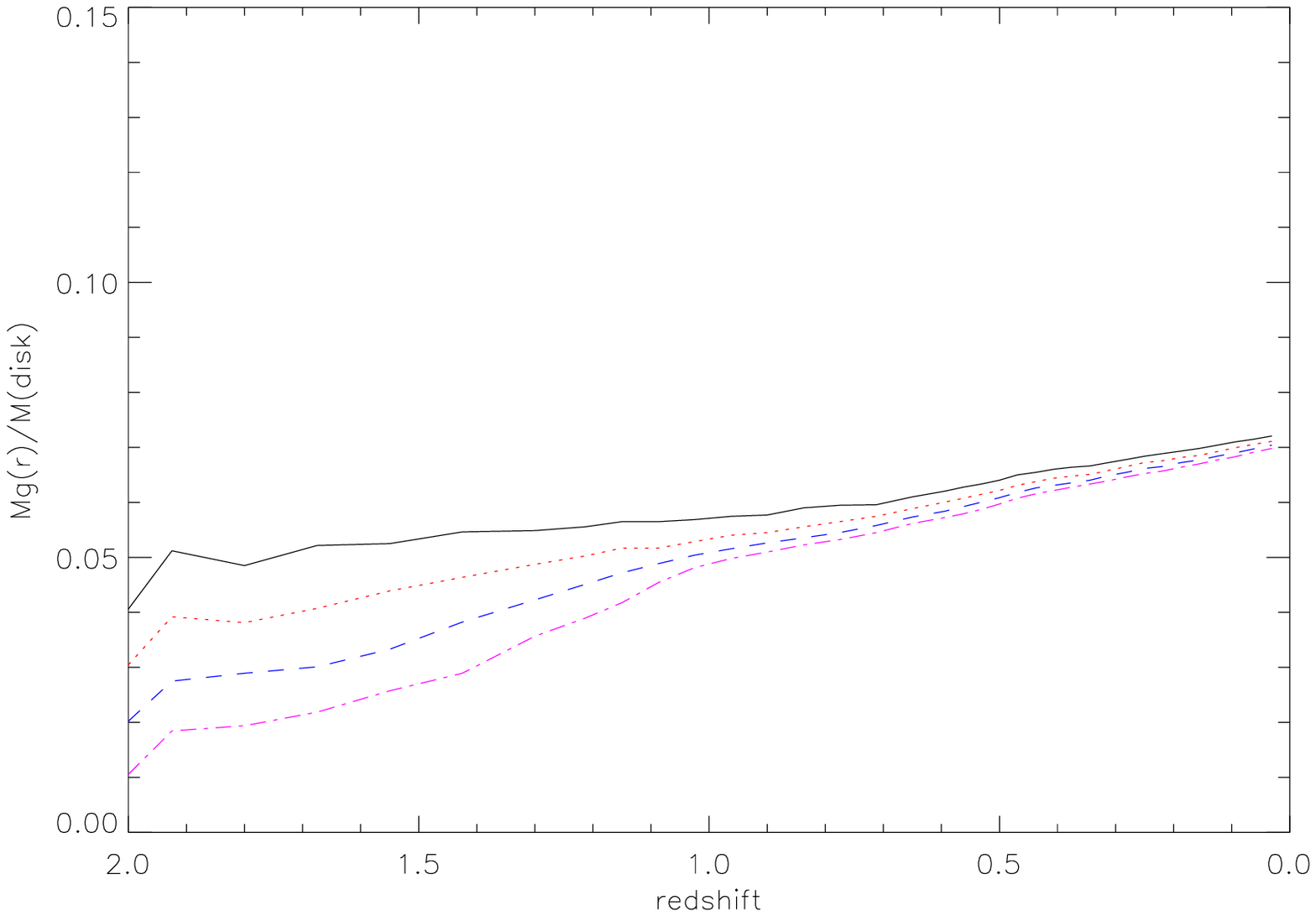}
\includegraphics[width=8cm]{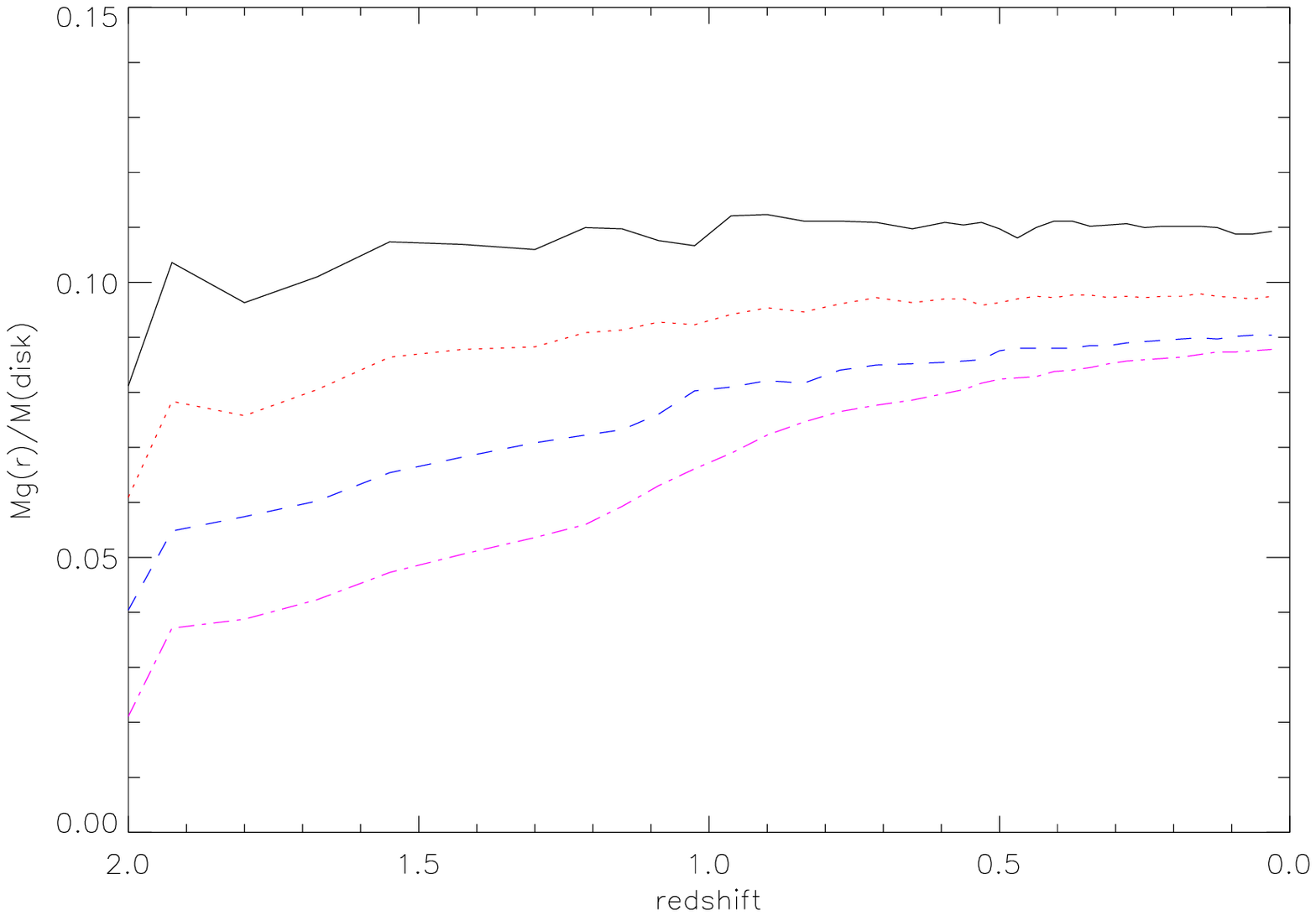}
\includegraphics[width=8cm]{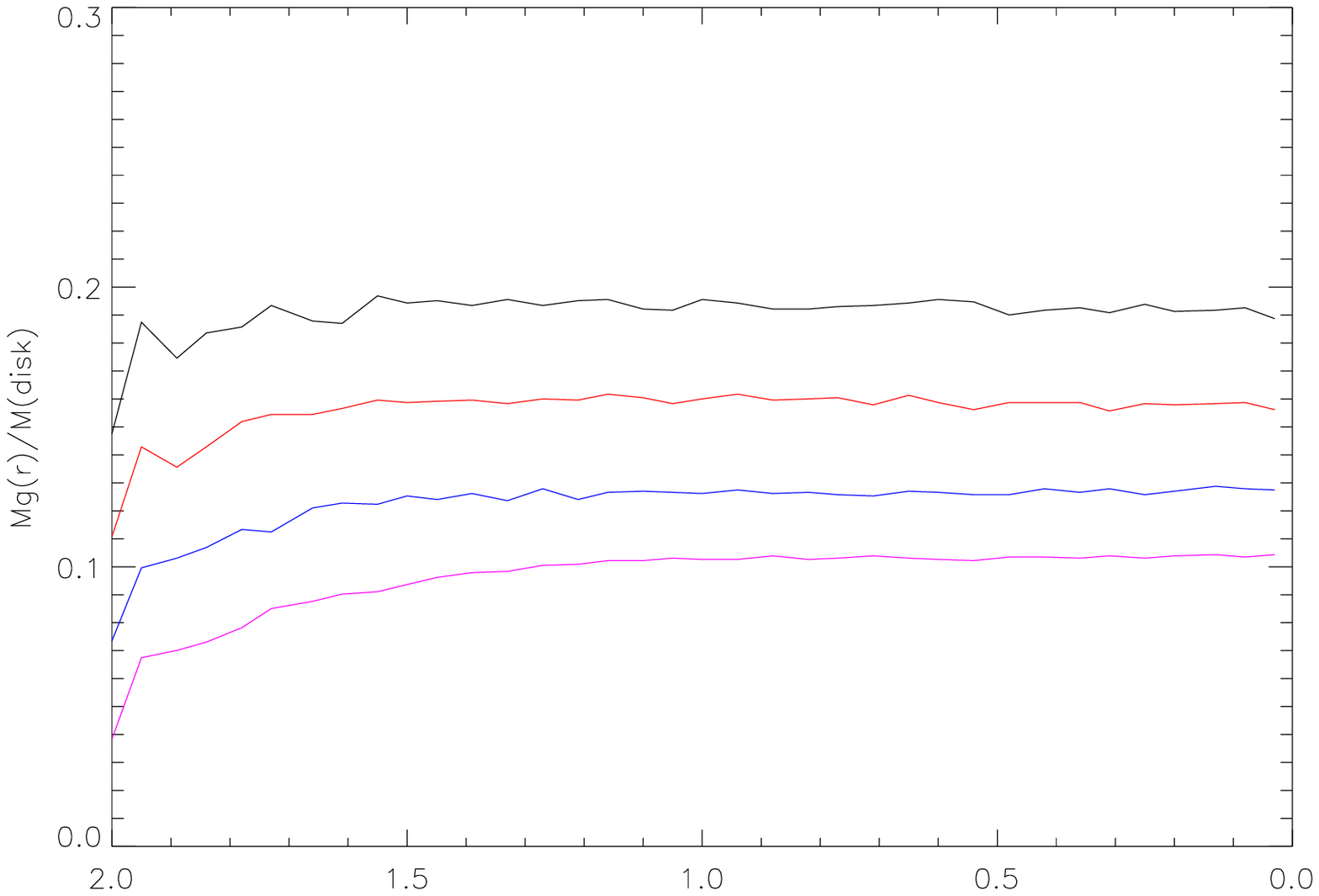}
\end{center}
\caption{As in Fig \ref{gas_m1} for simulation c1 (top panel), c2 
(middle panel), and c3 (bottom panel).}
\label{gas_m3}
\end{figure}

\begin{figure*}
\begin{center}
\includegraphics[width=8cm]{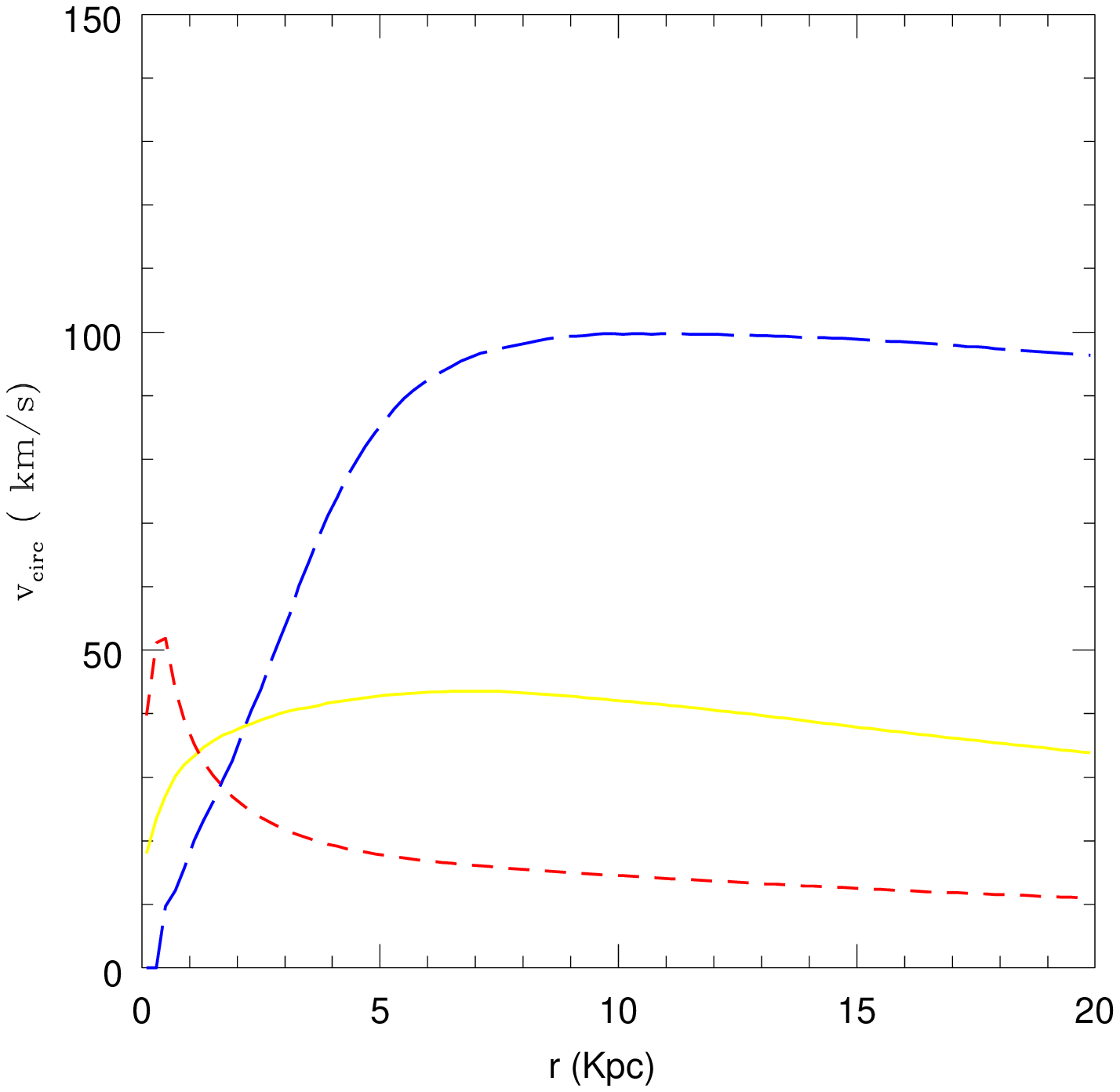}
\includegraphics[width=8cm]{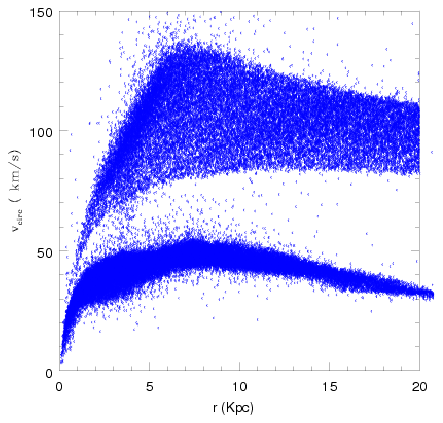}
\end{center}
\caption{ Left panel: circular velocities obtained assuming a spherical
  mass distribution of gas (red dashed line), stars (yellow
continuous line) and DM (blue long dashed line) at z=0 for simulation c4. Right panel: circular
velocities deduced from the radial force  (upper curve:
dark matter, lower curve: stars) at z=0 for
simulation c4.}
\label{vcirc_01_01}
\end{figure*}

\begin{figure*}
\begin{center}
\includegraphics[width=8cm]{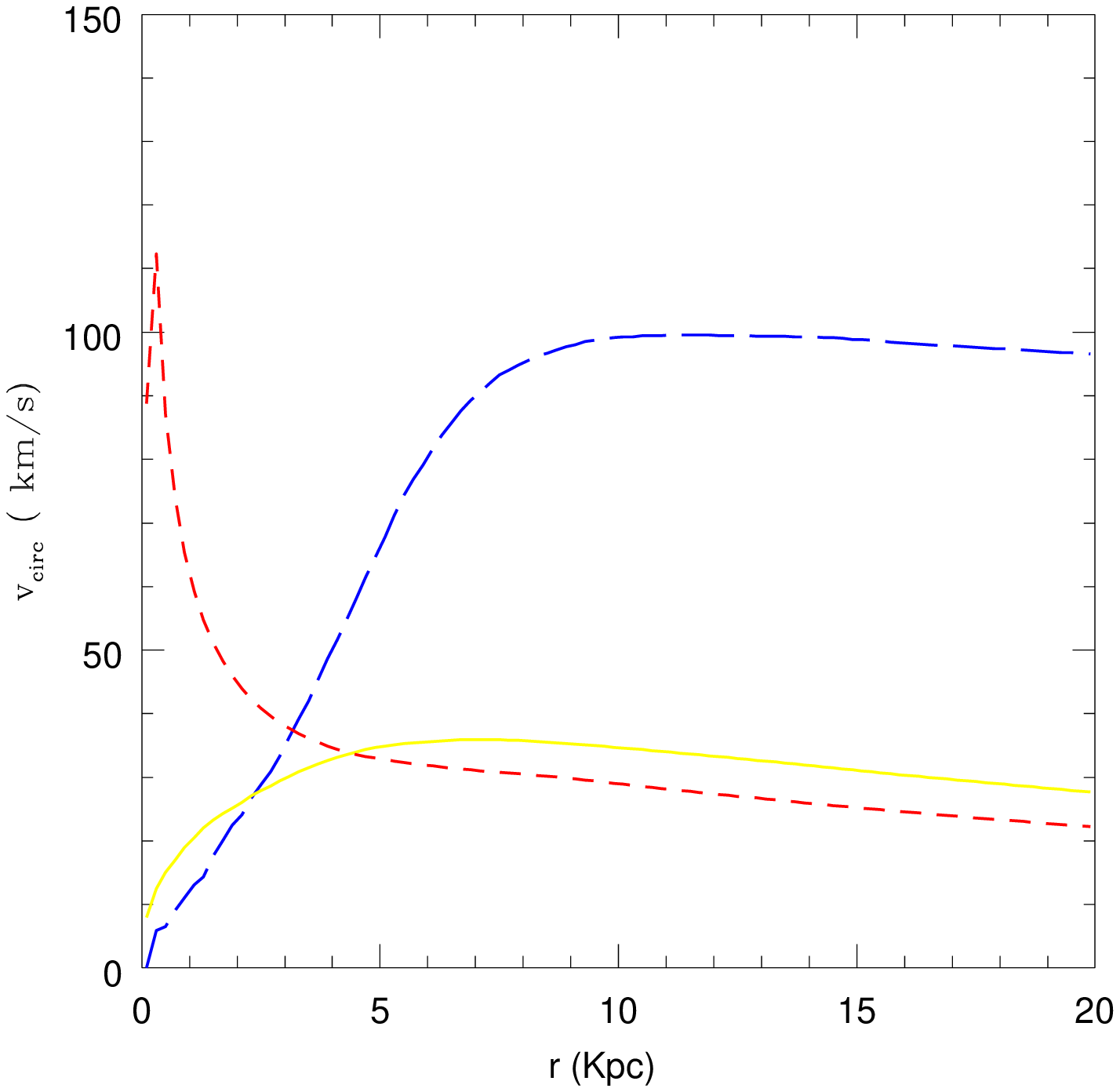}
\includegraphics[width=8cm]{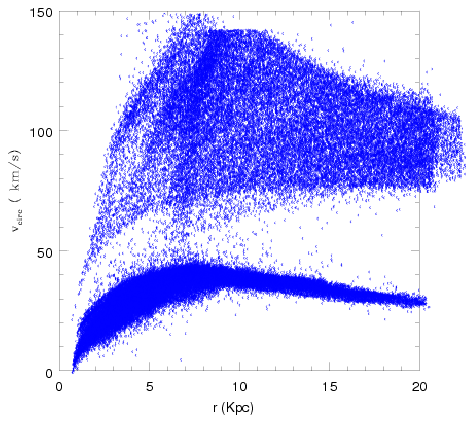}
\end{center}
\caption{ The circular velocities as in Fig.  \ref{vcirc_01_01} for simulation c6.}
\label{vcirc_01_04}
\end{figure*}

One crucial parameter to monitorate the ability of the central gas
condensation to destroy the bar, is indeed the ratio of the central gas 
mass over the total disk mass, that
we show in Fig. \ref{gas_m1} and \ref{gas_m3}.
 Fig. \ref{gas_m1}, for DM dominated disk simulations, 
show that the rate of inflow at each radius decreases for higher gas fractions:
from 65\% to 37\% at 2 Kpc
and from 56\% to 35\% at 5 Kpc as we go from simulation c4 to c8.
Fig. \ref{gas_m3}, for our more massive disks,
show the same trend but with a stronger decrease:
from  75\% to 35\% at  2 Kpc and from 66\% to 41\% at 5 Kpc 
as we go from simulation c1 to c2.
In particular, for simulation c2, where the gas fraction is able to destroy the bar, 
the accretion rate  at the outer radii  becomes quasi stationary after $z=1$ 
(Fig. \ref{gas_m3}, middle panel) since
with the   weakening 
of the bar, the gas inflow is no longer forced.
In simulation c3, such a behaviour appears
earlier (Fig. \ref{gas_m3}, middle panel) as well as the bar disappears earlier than in the c2 case. \\
 The  circular velocity,  v$_{circ}$, computed assuming a spherical mass
  distribution and therefore having the expression $(GM/r)^{1/2}$ where M is the mass of a
component of the system, gives  a measure of the absolute concentration of
such a component. 
 In  Figs.  \ref{vcirc_01_01} and  \ref{vcirc_01_04} (left panels)  we
 show its behaviour at $z=0$ for  gas, stars and dark matter in 
 simulations  c4 and c6. 
 In the same Figs. (right panels) we also show the behaviour of the
  v$_{circ}= r({\partial \Phi}/{\partial r}) $  evaluated
  from the force field of  the stellar and DM components.
  The difference  of these values  from  the previous   ones
    is due to the deviation from the spherical isotropy of the mass
    distribution. In particular, the wide spread we observe for the DM
    component indicates a wide anisotropy, wider at large radii.\\

The impact of a central mass
concentration (CMC) on the bar evolution has been discussed in detail  by
\citet{she04}.
Their study is not directly comparable with ours, since the growth of the CMC 
is modelled with an analytic potential instead of being  the result of  
gas inflow like in our
simulations. In  their work it was  shown that the destruction power
 depends on the CMC, given as 
{\it a fraction of the disk mass},  and on the 'softening' of its
potential,  r$_s$, a parameter which controls the compactness of the CMC.  
 \citet{she04}  find that for a {\it hard} CMC (namely with  $r_s$ very
small) only few percent of  the disk mass is able to
destroy the bar, whereas for a {\it soft} CMC ( $r_s =$ 0.3 Kpc) a fraction
 larger than   0.1 plays the job.
In our scenario, r$_s$  qualitatively
translates into  the radius, r$_g$, where we set the evaluation of  
the gas mass concentration.
We deduce that  for our more
massive disks (i.e.  when  R$_{DM}=1.94$ in Table 1)
a gas mass greater than 9\% of the disk 
mass inside $r_g = 2$ Kpc  is needed to destroy the bar. 
However for DM dominated disks,
 (i.e.  with R$_{DM}=6.4 $ in Table 1)  a value of  18\% of the disk
mass in the same r$_g$ cannot destroy the bar.

Therefore, our cosmological simulations point out that   also the
value of  $R_{DM} $  is a crucial parameter in this play.


\subsection{Bar pattern speed}
\begin{figure*}
\begin{center}
\includegraphics[width=8cm]{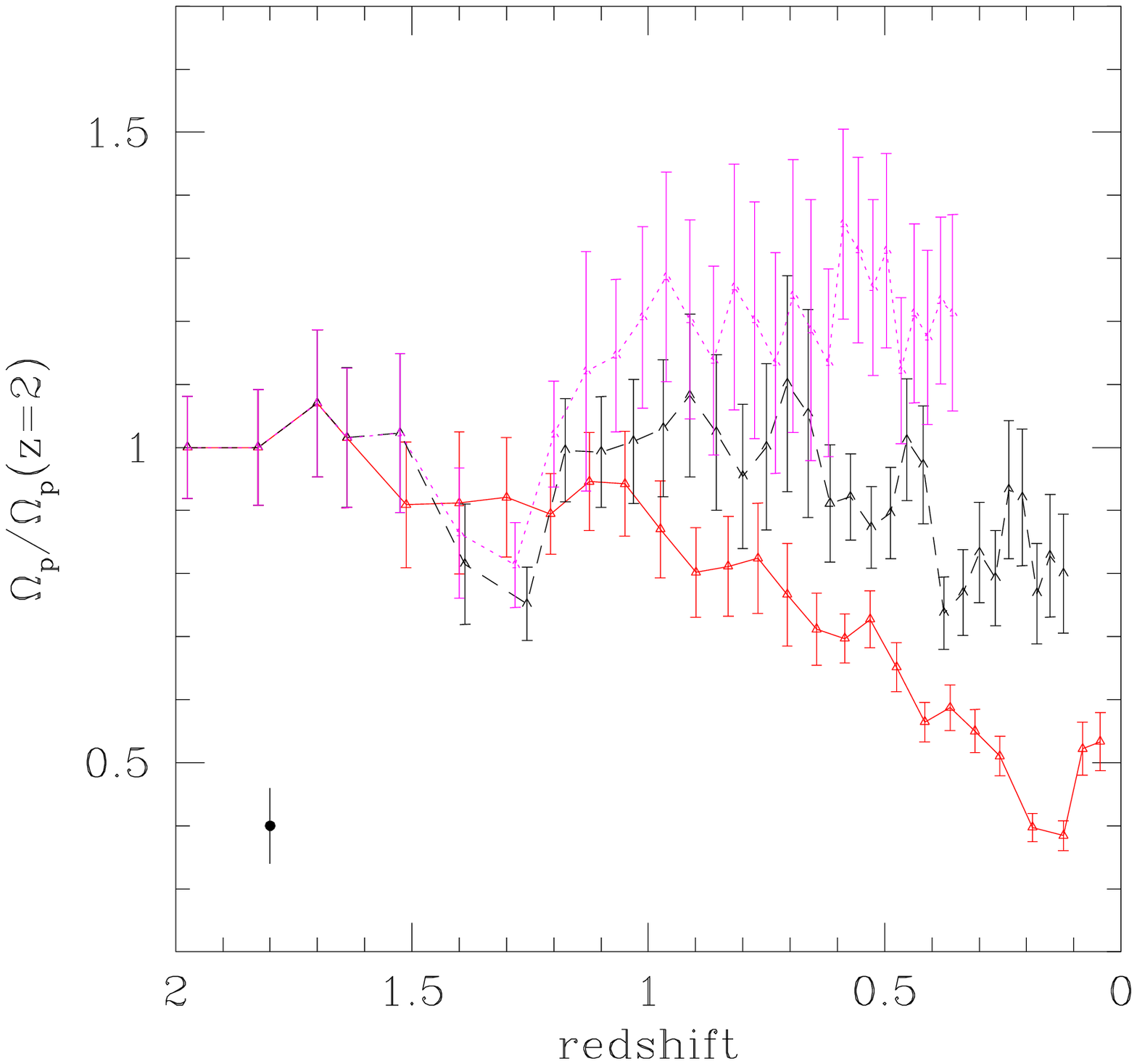}
\includegraphics[width=8cm]{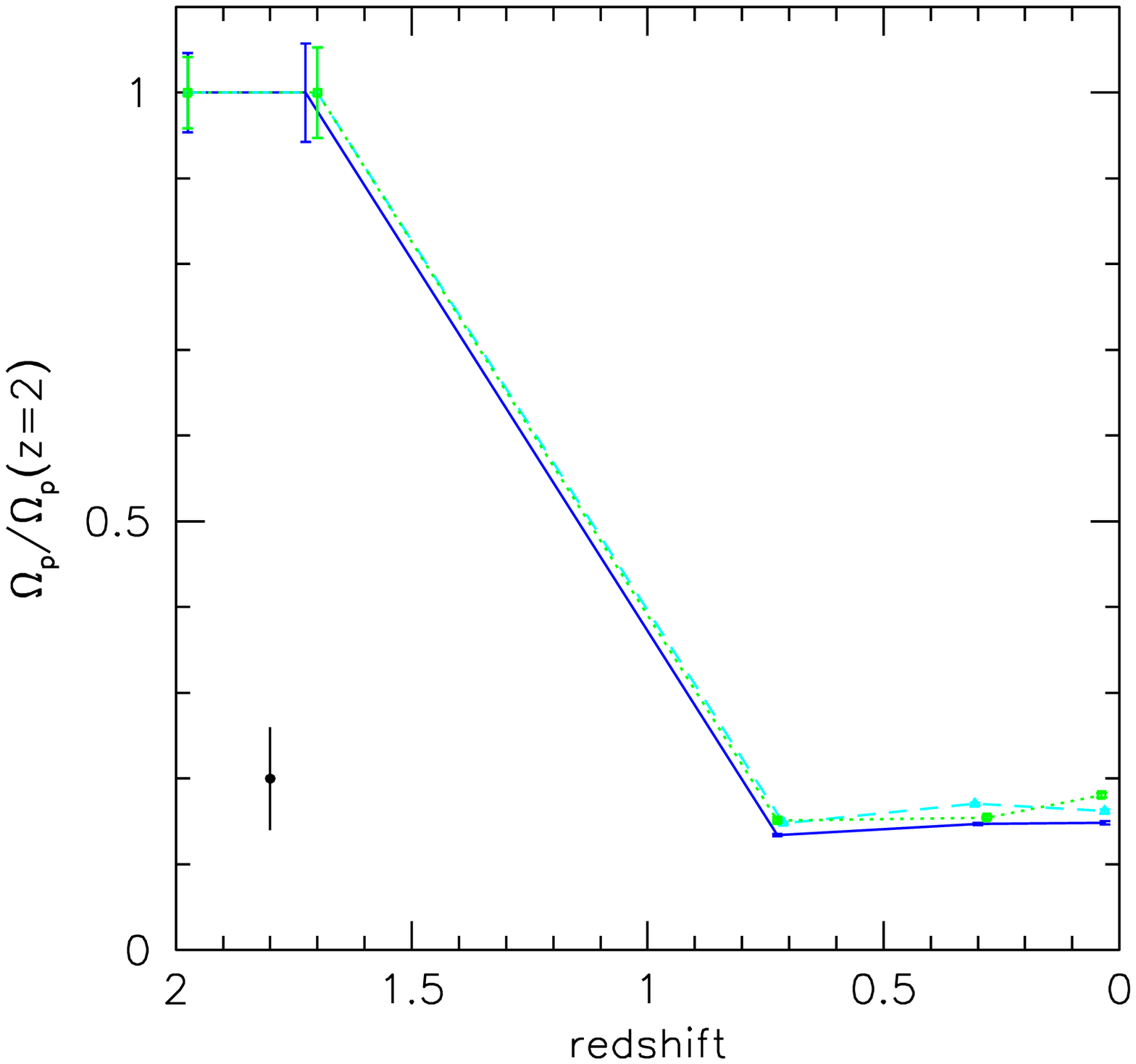}
\end{center}
\caption{ Evolution of the bar pattern speed with the redshift. Left panel:  simulation c1
  (red full line), c2 (black dashed line), c3 (magenta dotted line). Right
  panel: simulations c4 (blue
  full line), c5 (light blue dotted line), and  c6 (green  dashed line). The
  error bars on the curves are obtained with the running average method, and
  the error bar at the bottom of each panel represents the uncertainty in the
  measure of the position angle of the bar (see text). }
\label{pattern1}
\end{figure*}

The bar pattern speed, $ \Omega_p$ is the angular speed  of the
the bar-like  density wave as viewed from an inertial 
frame.  Here it is evaluated following the position angle of the major bar
axis during the  disk evolution, where the major axis is selected with the
same criterion E1 quoted in \citet{MiWo}, namely as the radius where the
ellipticity profile reaches a maximum.   In  Fig. \ref{pattern1}  the values of  $ \Omega_p$ as function of the redshift
  are shown for the two sets of our simulations: the more massive disks  and the lighter disks respectively. Since the
  behaviour of    $ \Omega_p$  is rather noisy, we provided also the error
  bars calculated with the running average method, i.e. they are obtained
  averaging three consecutive values and computing  their standard
  deviations.  This error bar is more representative
of the fluctuations of the quantity $ \Omega_p$  than of the real errors on  the
measurements. On the other hand,
  the uncertainty in the measure of the position angle of the
  semimajor axis is always less than  $10^o$. Such a limit  leads to a maximum
  error of  6\% in the  evaluation of $ \Omega_p$. The error bar related to
  such a maximum error is presented at the bottom of each panel in Fig.14.
For the more massive disks,  the stronger decrease  is 
observed in simulation c1. This is also the simulation with the 
longer and the stronger  stellar bar (see Table \ref{cosmsimtable_fin}), usually coupled with the higher decrease 
of the pattern speed during the evolution \citep{atha02}.
Simulation c2 shows the same trend,   
the decrease of $ \Omega_p$ is slightly lower, and the  bar disappears before 
z$=0$, as in simulation c3.  However in such a simulation,
the pattern speed increases by decreasing 
the redshift. 

We derive that  in the more
  massive disks  the  increasing gas fraction  produces a
speed up of the stellar bar pattern  since the
gas presence both shortens the bar and decreases its ellipticity. 
Moreover  such a  behaviour is consistent with
the findings of \citet{ber98}, i.e. an accelerated evolution of the whole  system when the gas
is  included.

On the other hand,  the DM dominated disks  show  the strongest decrease of  $
\Omega_p$  with the redshift.
%
 The lower dynamical support of these light  disks  compared
to the more massive cases indeed enhances the role of the gas which slows down the
pattern speed more quickly.

\section{Discussion and conclusions}

We presented eight  cosmological simulations  with the same disk--to--halo mass
ratios as  in  \citet{Cu06}. In order to study the impact of the gaseous 
component, here we included and 
varied its percentage inside disks with different disk--to--halo mass ratios.

Between the more massive disks, simulation
c1, with the smaller gas percentage, shows  a long lasting bar, 10 Gyr old, 
stronger
than the one developed in the pure stellar case. 
In simulation c2, a gas percentage of 20\% 
appears as  a threshold value for the bar life as far as such
disk mass is concerned. In such a simulation, indeed,  
the gas is able to destroy the bar at the very
end of the  evolution (z=0.15).  By increasing the gas percentage, as in  simulation c3,
the bar disappears earlier than in c2 (at z=0.6).
We find  that a
gaseous mass concentration equal to 9\% of the total mass of the disk
inside a radius of 2 Kpc, is a lower limit  for the bar dissolution. 
 In our more massive disks, where the baryons gravitational field is compelling with that 
of the DM halo (i.e. disk-to-halo mass ratio 0.33),
we find a threshold value of the gas fraction, 0.2, able to destroy the bar.\\ 
 
 The gas shows a trend to be more widely distributed 
by increasing its fraction, for a given disk mass. Moreover, for the same 
gas fraction,  the  
mass inflow is larger in more massive  disks. 
These effects contribute to
weakening the bar which, however, in our cosmological framework,
cannot be destroyed in DM dominated disks. In these cases, indeed, the cosmological
evolution of the DM halo drives the growth of the bar instability so that
the presence of the gas appears as a second order effect on its evolution. 
 
The  DM
dominated disks show indeed a behaviour which is strongly driven by the cosmology as 
in \citet{Cu06}.
In these cases : i) we do not find any  value of 
the gas fraction, in the range 0.1--0.6, able to destroy the bar, 
ii)  even a high value of central gas concentration
does not succeed in dissolving the  bar.  
The bars in
these disks are not a classical product of the self--gravity, which is very
weak, or of  angular momentum exchanges,
since the disk rotates very slowly, but they are
features that strongly  depend on  the dynamical state and evolution of the cosmological halo. 
Therefore the classical
results emphasising the gas impact obtained outside the cosmological scenario are
no longer applicable. 
This confirms the results of \citet{Cu06}, where 
 in the DM dominated disks
the bar feature is triggered and maintained by the cosmological
characteristics of the halo, namely its triaxiality and its dynamical state. 

Therefore the whole set of cosmological simulations we present here is 
suggesting a play between three parameters which drive the bar formation and
dissolution in cosmology:  the  gas fraction inside a suitable disk radius, 
$r_g$,  the $r_g$ value itself, and the halo-to-disk
mass ratio inside the disk radius.

{\bf Acknowledgements}  
Simulations have been performed on the CINECA IBM SP4 computer, thanks
to the INAF-CINECA grants cnato43a/inato003 ``Evolution of disk
galaxies in cosmological contexts'', and on the Linux PC Cluster of
the Osservatorio Astronomico di Torino. We wish to thank for useful
discussions:  L. Athanassoula, I. Berentzen,  F. Bourneaud, E. D'Onghia,
A. Klypin \& V. Springel.

\end{document}